\documentclass[prl,twocolumn,superscriptaddress,noshowpacs,notitlepage,longbibliography]{revtex4-1}
\usepackage{graphicx,bm,times}
\usepackage{amsmath}
\usepackage{amsfonts}
\usepackage{amssymb}
\usepackage{color}
\usepackage{hyperref}
\hypersetup{
	colorlinks = true,
	allcolors = {blue}
}

\begin{document}
\title{On the nature of quasiparticle interference in three dimensions}
\author{Luke C. Rhodes}
\affiliation{SUPA, School of Physics and Astronomy, University of St Andrews, North Haugh, St Andrews, Fife, KY16 9SS, United Kingdom}
\author{Weronika Osmolska}
\affiliation{SUPA, School of Physics and Astronomy, University of St Andrews, North Haugh, St Andrews, Fife, KY16 9SS, United Kingdom}
\author{Carolina A. Marques}
\affiliation{SUPA, School of Physics and Astronomy, University of St Andrews, North Haugh, St Andrews, Fife, KY16 9SS, United Kingdom}
\author{Peter Wahl}
\affiliation{SUPA, School of Physics and Astronomy, University of St Andrews, North Haugh, St Andrews, Fife, KY16 9SS, United Kingdom}
\date{November 2021}

\begin{abstract}
Quasiparticle Interference (QPI) imaging is a powerful tool for the study of the low energy electronic structure of quantum materials. However, the measurement of QPI by scanning tunneling microscopy (STM) is restricted to surfaces and is thus inherently constrained to two dimensions. This has proved immensely successful for the study of materials that exhibit a quasi-two-dimensional electronic structure, yet it raises questions about how to interpret QPI in materials that have a highly three dimensional electronic structure. In this paper we address this question and establish the methodology required to simulate and understand QPI arising from three dimensional systems as measured by STM. We calculate the continuum surface Green's function in the presence of a defect, which captures the role of the surface and the vacuum decay of the wave functions. We find that defects at different depths from the surface will produce unique sets of scattering vectors for three dimensional systems, which nevertheless can be related to the three-dimensional electronic structure of the bulk material. We illustrate the consequences that the three-dimensionality of the electronic structure has on the measured QPI for a simple cubic nearest-neighbour tight-binding model, and then demonstrate application to a real material using a realistic model for PbS. Our method unlocks the use of QPI imaging for the study of quantum materials with three dimensional electronic structures and introduces a framework to generically account for $k_z$-dispersions within QPI simulations.
\end{abstract}

\maketitle

Quasiparticle interference (QPI), the spatial perturbation to the local density of states (LDOS) in the presence of defects or boundaries, is an important phenomenon that enables Scanning Tunneling Microscopy (STM), a real-space technique, to uncover information about electronic states in momentum space\cite{Crommie1993,hasegawa_direct_1993} with unparalleled temperature and energy resolution \cite{ast_sensing_2016}. The perturbations due to QPI arise as a direct result of scattering between two electronic states and therefore measuring the Fourier transform of these spatial perturbations provides a route to uncover the electronic structure of a material \cite{Petersen1998,Simon_2011}.


These measurements have been immensely successful in understanding materials with a highly 2D electronic structure, such as the cuprates\cite{Hoffman2002,kohsaka_how_2008,He2014_Cuprate} and ruthenates \cite{wang_quasiparticle_2017,Kreisel2021}, where a comparison between experimental measurements and theoretical models is rather straightforward. On the other hand, in anisotropic materials with non-negligible interlayer hopping, such as the iron-based superconductors \cite{allan_anisotropic_2012,allan_identifying_2014,Sprau2017} or heavy Fermion systems \cite{schmidt_imaging_2010,zhou_visualizing_2013,Akbari2014}, it has become apparent that the resulting three-dimensionality of the electronic structure results in visible changes to the measured QPI beyond a simple 2D model\cite{Hanaguri2018, Rhodes2019}.

The challenge to understand QPI of 3D electronic structures is fundamentally linked to the fact that STM is a surface-sensitive technique, limited to measuring the LDOS at surfaces and in two spatial dimensions. This raises an important question about how to interpret Fourier transforms of QPI measurements in systems that have a notable three-dimensional electronic structure.

Previous theoretical and experimental work studying the QPI of 3D electronic structures have shown that the direction and intensity of the QPI standing wave patterns are controlled by the Fermi velocity of the electronic states \cite{Weismann2009,Lounis2011,Kotzott_2021}. More recently, in conjunction with the experimental constraint that STM is a local technique, it has been argued that any standing waves generated by electronic states which have finite group velocity in the Z direction \cite{Lounis2011}, will actually traverse into the bulk of the material and therefore not generate coherent long range QPI signal that would be noticeable as sharp peaks in the Fourier transform \cite{Rhodes2019,marques_tomographic_2021}. 

So far however, these arguments have not considered the role that the surface has on the electronic states, where $k_z$ is no longer a good quantum number. In this letter, we address this issue by theoretically studying the consequence that the surface has on the formation of QPI. By utilising the recently developed continuum LDOS (cLDOS) technique\cite{choubey_visualization_2014,Kreisel2015}, 
which takes into account the inter-unit cell superposition of the electronic states above the surface of a material, we show that information about the bulk 3D electronic structure can be readily obtained from experimental QPI measurements. We additionally show that defects at different distances from the surface will produce unique QPI patterns and that a full comparison between theory and experiment requires the consideration of defects at both different sites and depths from the surface. 


To simulate QPI, we begin by calculating the cLDOS $\rho(\mathbf{r},\omega)$ using the continuum Green's function $G(\mathbf{r},\mathbf{r},\omega)$ following  Ref.~\onlinecite{choubey_visualization_2014,Kreisel2015,Kreisel2021}

\begin{equation}
    \rho(\mathbf{r},\omega) = -\frac{1}{\pi}\mathrm{Im} G(\mathbf{r},\mathbf{r},\omega).
    \label{Eq:LDOS}
\end{equation}

Here, $\mathbf{r}$ is a 3D continuous real space vector, $\omega$ is the energy and the continuum Green's function is defined via the continuum transformation, using Gaussian-type orbitals, of the discrete lattice Green's function that encorporates a point-like defect. Details can be found in the supplemental material. 


The cLDOS, $\rho(\mathbf{r},\omega)$, is then calculated on a large real-space grid over $r_x$ and $r_y$ at a fixed height $r_z=h$,  and the 2D fourier transform is taken to generate $\tilde{\rho}(\mathbf{q_\parallel},\omega)$ which can be compared with experimental measurements.

To understand the influence of the 3D electronic structure on QPI measurements, we begin by considering the simple cubic lattice with nearest neighbour hopping 

\begin{equation}
    H(\mathbf{k}) = t_x\cos(k_x) + t_y\cos(k_y) + t_z\cos(k_z).
    \label{Eq:3D_ham}
\end{equation}

\begin{figure}
    \begin{center}
    \includegraphics[width=\linewidth]{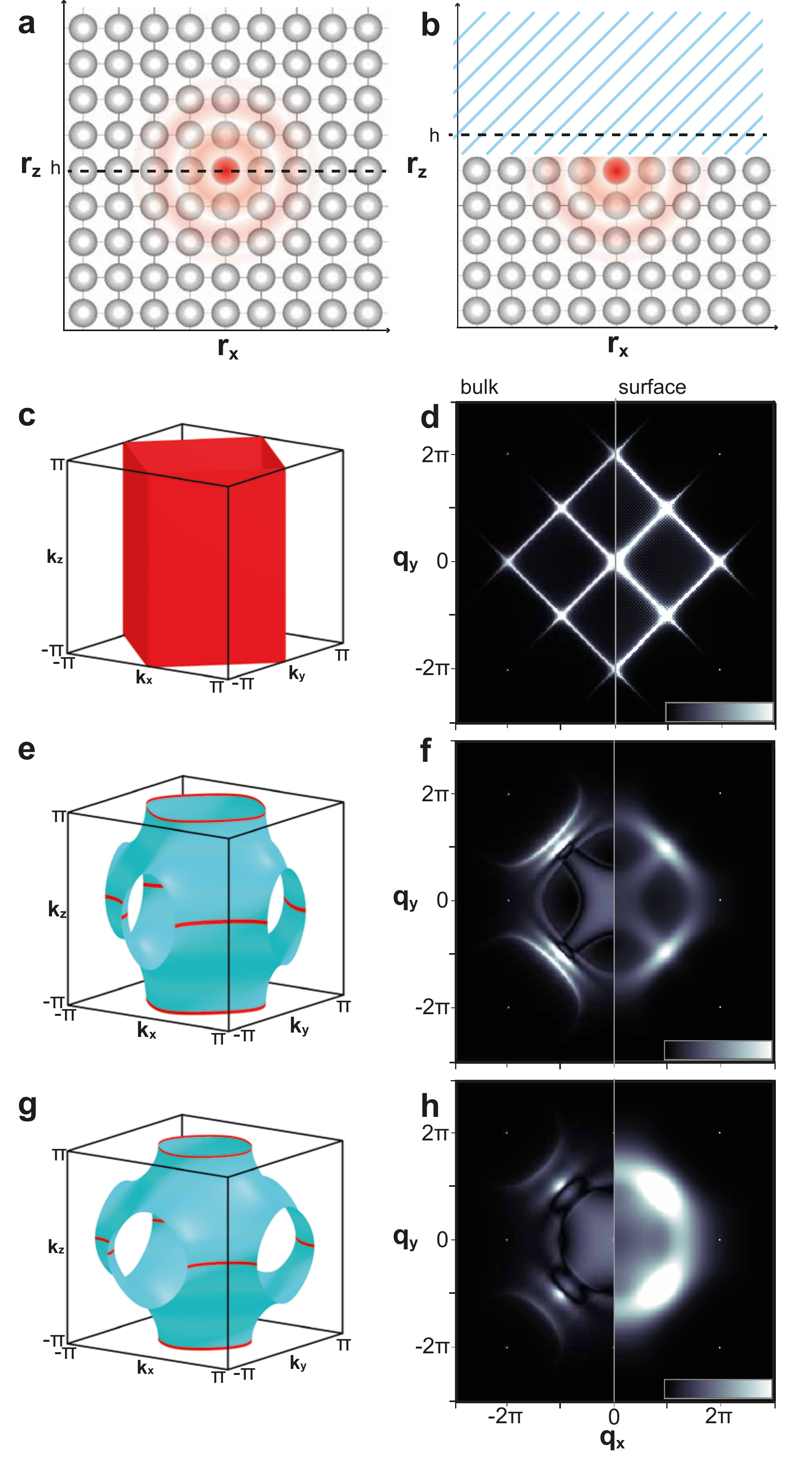}
    \end{center}
    \caption{\textbf{Consequence of out of plane hopping on the bulk and surface QPI patterns}. a, b) sketch of the two scenarios considered, a) a defect situated in the bulk of a material, b) a defect situated at the surface of a material. c, d) 3D Fermi surface and cQPI, $\tilde{\rho}(\mathbf{q_\parallel},\omega)$, calculated for the bulk defect (left half) and surface defect (right half) for the nearest neighbour cubic model with $t_x = t_y = 0.1$~eV and $t_z= 0$~eV. e, f) Equivalent simulations for $t_z = 0.05$~eV, g, h) $t_z = 0.1$~eV (isotropic case).}
    \label{Fig:fig1}
\end{figure}

We then consider two scenarios, one where the defect is located deep in the bulk of the material, as sketched in Fig.~\ref{Fig:fig1}(a), and the second for a defect located at the surface of a large slab, as sketched in Fig.~\ref{Fig:fig1}(b). For the former we calculate Eq.~\eqref{Eq:LDOS} using the Hamiltonian in Eq.~\eqref{Eq:3D_ham}, whereas for the surface calculation we first perform a co-ordinate transformation to an $N$-layered slab,


\begin{equation}
H(\mathbf{k_\parallel}) = 
\begin{pmatrix}
   H^{0}(\mathbf{k_\parallel}) & H^{1}(\mathbf{k_\parallel}) & H^{2}(\mathbf{k_\parallel}) &  ... \\
   H^{1}(\mathbf{k_\parallel}) & H^{0}(\mathbf{k_\parallel}) & H^{1}(\mathbf{k_\parallel}) &   ... \\
     H^{2}(\mathbf{k_\parallel}) & H^{1}(\mathbf{k_\parallel}) &H^{0}(\mathbf{k_\parallel}) & ... \\
     ... & ... & ... & ... \\
    \end{pmatrix},
\label{Eq:Ham_Slab}
\end{equation}

where we have separated the Hamiltonian such that each row and column in the $N\times N$ block matrix describes a one-unit-cell thick layer along the $z$ axis and the crystallographic momentum is now defined parallel to the XY plane ($\mathbf{k_\parallel} = (k_x,k_y)$). The individual elements are then defined as

\begin{equation}
    H^{R_z}(\mathbf{k_\parallel}) = \sum_{\mathbf{R_\parallel}} H(\mathbf{R_\parallel},R_z) e^{i\mathbf{k_\parallel}\mathbf{ R_\parallel}}. 
\end{equation}

For the nearest neighbour cubic model, $H(\pm 1,0,0) = t_x$, $H(0, \pm 1,0) = t_y$, $H(0, 0,\pm 1) = t_z$.

Without loss of generality, we assume that the surface does not distort the physical structure, nor induce any charge imbalance. To describe these would require modification of $H^0(\mathbf{k}_\parallel)$ for the surface layer(s) in Eq. \eqref{Eq:Ham_Slab}. 

\begin{figure*}
    \begin{center}
    \includegraphics[width=\linewidth]{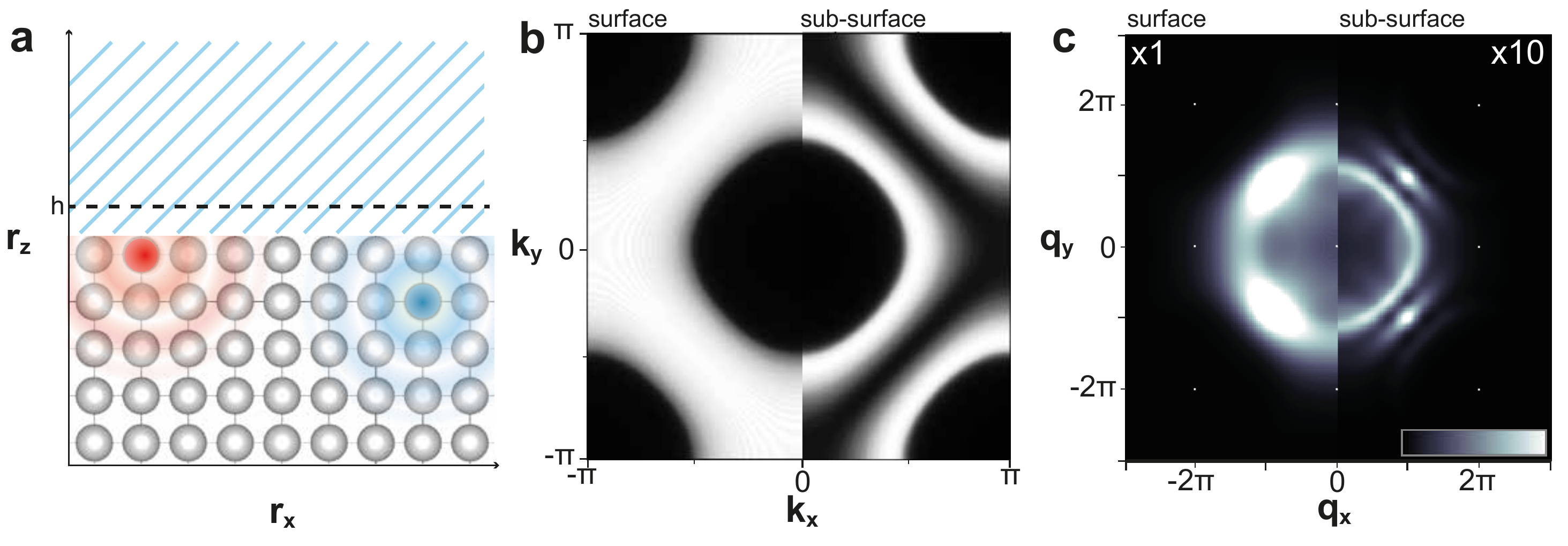}
    \end{center}
    \caption{\textbf{Consequence of defect position on measured QPI.}  a) Sketch of defects at different positions from the surface. b) Partial spectral function of the surface layer (left half) and sub-surface layer (right half). c) cQPI $\tilde{\rho}(\mathbf{q_\parallel},\omega)$ for a defect located at the surface (left half, red atom in (a)) and sub-surface (right half, blue atom in (a)). The intensity of the simulated cQPI due to a sub-surface defect has been enhanced by a factor of 10 for comparison with the surface cQPI pattern.}
    \label{Fig:fig2}
\end{figure*}


We begin by comparing the QPI pattern in the cLDOS (cQPI) at the Fermi level expected from a bulk or surface calculation when out-of-plane hopping is neglected ($t_z = 0$). We find that the two dimensional square Fermi surface (Fig.~\ref{Fig:fig1}(c)) will produce a square QPI pattern, regardless of whether a bulk or surface defect is considered. This is shown in the left and right hand sides of Fig.~\ref{Fig:fig1}(d) respectively. This is expected, as the slab Hamiltonian of Eq.~\eqref{Eq:Ham_Slab} will be diagonal in the absence of out of plane hopping ($H^1(\mathbf{k_\parallel})=0$). As $t_z$ is increased, however, the differences between a bulk and surface cQPI simulations become increasingly apparent.
Bulk simulations, shown on the left hand side of Fig.~\ref{Fig:fig1}(f, h), produce several sharp scattering vectors, which can be linked predominately to nesting of states with $k_z=0$ or $k_z = \pi$, shown as red lines in Fig.~\ref{Fig:fig1}(e, g), as well as some broader intensity emanating from the center due to poorly nested scattering between the  states with different $k_z$. Surface cQPI simulations do not produce such sharp scattering vectors. Whilst the qualitative features do resemble those observed for bulk calculations, the uncertainty in $k_z$ produces much broader scattering vectors and this broadness increases with increasing $t_z$. 

As all experimental measurements are performed above the surface of a material, this implies that measurements on materials with a highly three dimensional electronic structure will observe broad QPI scattering patterns from surface defects. Nevertheless, the presence of a surface also creates an additional consideration about the position of the defect. Whilst in a bulk material, defects in different unit cells will only induce a phase shift in the simulated cQPI, near the surface, the depth of the defect generates unique environments (Fig.~\ref{Fig:fig2}(a)). This will alter the possible scattering vectors that can be observed. To illustrate this, we plot in Fig.~\ref{Fig:fig2}(b) the partial spectral function $A^z(\mathbf{k}_\parallel,\omega)$ for the surface and sub-surface layers of the nearest neighbour cubic model with isotropic hopping,

\begin{equation}
A^z(\mathbf{k_\parallel,\omega}) = -\frac{1}{\pi} \mathrm{Im} G^{zz}(\mathbf{k_\parallel,\omega}),
\end{equation}

with $G^{zz}$ being the matrix element of the non-interacting Green's function for the $z$-th layer of the $N$-layer slab (see supplemental material). Here, we observe that the electronic states at the surface have a very different spectral density depending on whether we are looking at the surface unit cell (left hand side of Fig.~\ref{Fig:fig2}(b)) or the sub-surface unit cell (right hand side of Fig.~\ref{Fig:fig2}(b)). At the surface, the spectral weight is spread out over the entire range of $\mathbf{k_\parallel}$ spanned by the bulk 3D Fermi surface (Fig.~\ref{Fig:fig1}(g)) with a maximum intensity at $k_z = \frac{\pi}{2}$, whereas the sub-surface spectral function exhibits a suppression of spectral weight around $k_z = \frac{\pi}{2}$ and maximum intensity around $k_z =\frac{\pi}{4}$ and $\frac{3\pi}{4}$. It is also found that the projected spectral function at deeper layers will introduce additional nodes which eventually converge to the bulk $k_z$ averaged spectral function for large number of layers $N$ in the slab, as shown in Fig.S2 in the supplementary material.

\begin{figure*}
    \begin{center}
    \includegraphics[width=0.9\linewidth]{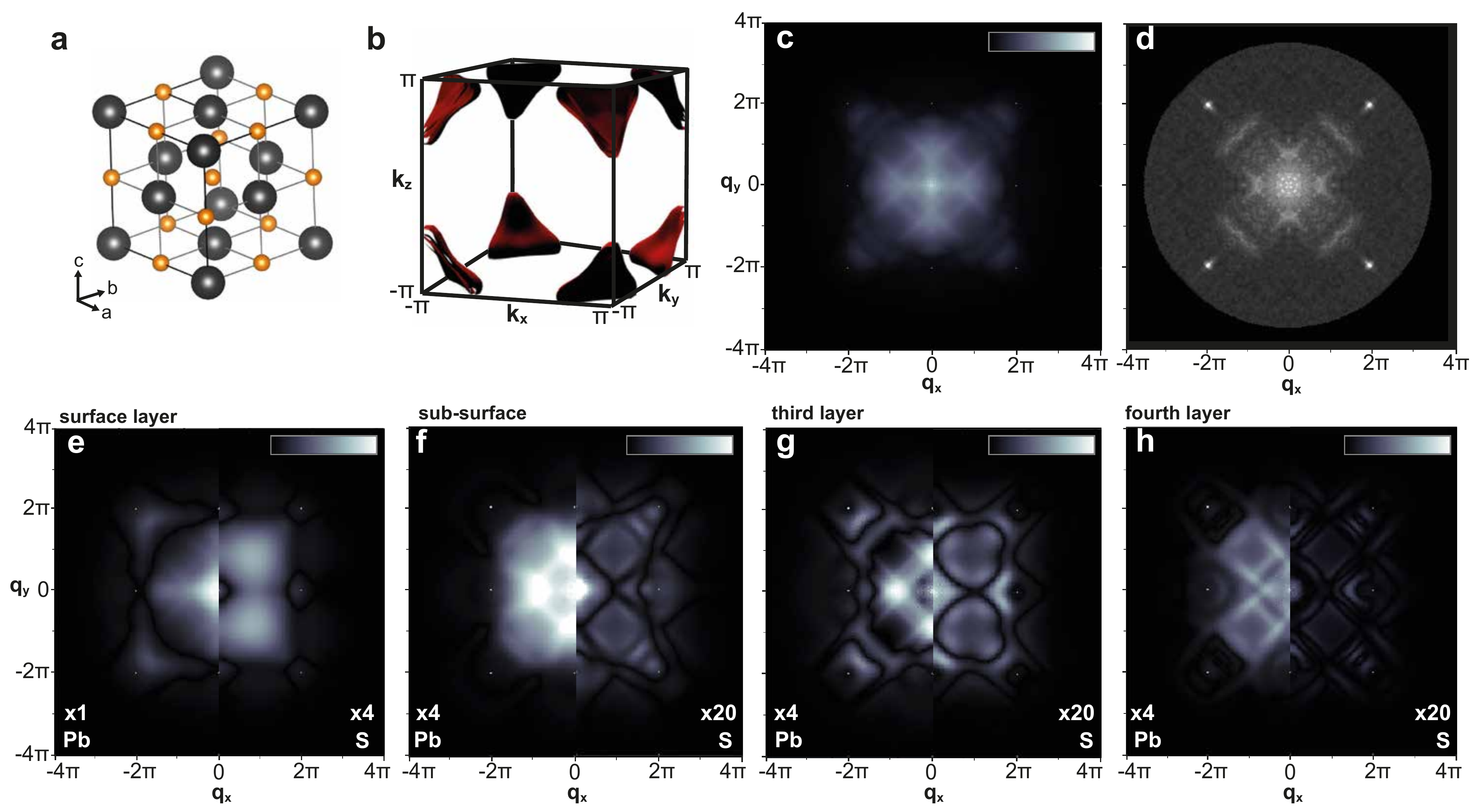}
    \end{center}
    \caption{\textbf{cQPI calculation for the rocksalt structure PbS.} a) Crystal structure of PbS, with larger black atoms as Pb and smaller orange atoms as S. b) 3D electronic structure taken at $E = 1$~eV above the Fermi level. c) Total QPI pattern obtained via calculating Eq.~\eqref{Eq:SumAll} of the main text with the weights $\alpha_i$ set to 1. The sum was performed over four Pb defects from the top four surface layers and four S defects from the top four surface layers. d) Experimental differential conductance QPI image of PbS from Ref. \cite{marques_tomographic_2021}, taken at $V = 0.80$~eV. The individual contributions of (c) are shown in Fig. (e-h) the Pb defect at a specific depth is shown on the left hand side of each panel and the S defect is shown on the right hand side. The maximum intensity of the colorscale has been defined relative to the left hand side of panel (e).}
    \label{Fig:fig3}
\end{figure*}

This effect is a consequence of quantum interference in the $z$-direction due to the surface breaking the translation symmetry resulting in resonator-like states in the vicinity of the surface, much like the quasi-particle interference of a one-dimensional defect in a 2D electron gas\cite{Crommie1993}. This has a pronounced effect on the spectral function, modulating the spectral density of the states in different depths and thus the strength with which scattering vectors for defects located at certain depths are observed. QPI scattering vectors arising from sub-surface and deeper defects will produce qualitatively different scattering vectors compared to surface defects, as shown in Fig.~\ref{Fig:fig2}(c) and these scattering vectors can be related to the full-three dimensional electronic structure of the bulk material by analysing the corresponding spectral function at a specific depth from the surface. For defects at a known depth, one may use this knowledge to extract information about the $k_z$ dispersion of the full three dimensional electronic structure, or, conversely, for a known electronic structure the scattering pattern can be used to determine the defect depth. We note that this is a general phenomenon resulting in a modulation of the spectral function $A^z(k_\parallel,\omega)$ as a function of depth $z$.

It is worth noting that the intensity of the QPI from sub-surface defects is rather weak. As illustrated in Fig. \ref{Fig:fig2}(c), the intensity of QPI arising from a sub-surface defect in this toy model is only 10\% of that from a surface defect. Nevertheless, this does not imply that the QPI from these defects can not be observed. In fact, it implies that if one wishes to directly reproduce experimental measurements of QPI arising from 3D systems, then one needs to sum the contribution of all types of defects, $i$, not just of different elements, but from different sites and depths, to the total cLDOS $(\rho^{i}(\mathbf{q},\omega))$ and multiply these by a realistic approximation for the different number of each type of defect

\begin{equation}
\rho^{\mathrm{exp}}(\mathbf{q_\parallel},\omega) =  \sum_i \alpha_i \rho^{i}(\mathbf{q_\parallel},\omega).
\label{Eq:SumAll}
\end{equation}

In some systems, it may be sufficient to only consider surface defects, however this will be dependent on the materials chemical composition and the relative number of each type of defect. For intrinsic bulk defects, $\alpha_i$ should be approximately equal and independent of $i$.

To illustrate this, in Fig.~\ref{Fig:fig3} we present equivalent cQPI slab calculations for a density functional theory derived tight binding model of PbS, a semiconducting rock-salt material shown in Fig.~\ref{Fig:fig3}(a) where QPI measurements have recently been reported \cite{marques_tomographic_2021}. The valence bands of this material are dominated by the $p$-orbitals of Pb and at an energy of 1~eV above the Fermi level exhibits a 3D electronic structure with no states around  $k_z = 0$, as shown in Fig.~\ref{Fig:fig3}(b). In Fig.~\ref{Fig:fig3}(c), we present the result of calculating Eq.~\eqref{Eq:SumAll} for a 16-unit cell thick slab of PbS, for the (100) surface, assuming all Pb and S defects are equally likely. This produces a complex pattern which is in good qualitative agreement with the experimental QPI measurement from Ref.~\cite{marques_tomographic_2021} at a similar energy (0.8 eV) shown in Fig.~\ref{Fig:fig3}(d). 

Fig.~\ref{Fig:fig3}(e-h) reveals the power of the methodology employed here. Each Pb defect (left) and S defect (right) that was considered produces a unique QPI scattering pattern depending on the relative distance from the surface, and it can be seen that some scattering patterns (e.g. left hand side in Fig.~\ref{Fig:fig3}(f) and \ref{Fig:fig3}(h)), are remarkably similar to the experimental measurement. This could suggest an uneven distribution of $\alpha_i$ in this system Ref.~\cite{marques_tomographic_2021} and highlights that this numerical simulation technique can be used to identify not only the three dimensional electronic structure of a system but also the relative concentration of unique defects at or below the surface, assuming the electronic structure is sufficiently well understood.

Our analysis of cQPI simulations for three dimensional systems highlights an important consideration regarding the comparison and interpretation of QPI measurements. Defects of the same type, but at different distances from the surface will produce unique sets of scattering vectors governed by unique regions of $k_z$. It implies that if one wishes to truly compare and understand the electronic structure of 3D materials using QPI, one needs to compare the experimental measurement with a simulation that takes into account multiple defects not just of different types, but at different depths from the surface. Additionally, unless 2D surface states are present \cite{Lambert2017,Pinon2020,Russmann2021}, if the system exhibits non-negligible out-of-plane hopping, the sharpest features present in Fourier transformed QPI measurements will originate not from defects at the surface but from sub-surface defects.

These details were not captured in previous theoretical studies of three dimensional systems \cite{Derry2015,Lambert2017,Rhodes2019,marques_tomographic_2021}, due to the use of discrete, site-centered, Greens functions. The continuum transformation employed here is therefore a very useful tool to reliably and accurately compare STM measurements with theoretical simulations for generic materials. 

In this work, we performed this continuum transformation assuming Gaussian-or Slater-type orbitals, and left the radii of these orbital as a free parameter which could be fit to experimental measurements (See supplemental material), however it is also possible to more accurately capture the orbital overlap and decay into vacuum by explicitly calculating the Wannier orbitals, e.g from density functional theory \cite{choubey_visualization_2014,Kreisel2015,Kreisel2021}. This may be required in more complex systems with multiple atomic elements, particularly in systems where the surface layer does not contribute any states to the Fermi level.

To summarise, we have studied how out-of-plane hopping modifies the electronic response in realistic simulations of QPI. Our results provide a generic framework to understand this behaviour, and highlight the importance of defect depth, position and type on experimental observables. The methods introduced here unlock the ability to understand the three dimensional electronic structure of materials using scanning tunneling microscopy. The formalism presented here can be used analogously to describe the behaviour of magnons \cite{mitra_magnon_2021} or phonons near surfaces, providing a broader framework to describe how quasi-particles can be detected at surfaces.

\begin{acknowledgments} 
LCR acknowledges support through the Royal Commission for the Exhibition 1851 and CAM and PW from the Engineering and Physical Sciences Research Council (EPSRC EP/L015110/1, EP/S005005/1 and EP/R031924/1). Underpinning data will be made available at \cite{3dqpidata}. For the purpose of open access, the authors have applied a Creative Commons Attribution (CC BY) licence to any Author Accepted Manuscript version arising.
\end{acknowledgments}

\appendix
\begin{widetext}
\section{cQPI simulations}

To simulate the Quasiparticle Interference (QPI), we start with the discrete lattice Greens function $G(\mathbf{R},\mathbf{R^\prime},\omega)$ in the presence of a point-like impurity with impurity potential $\hat{V}$. The lattice Green's function in presence of the defect is obtained from the unperturbed lattice Green's function $G_0(\mathbf{R},\mathbf{R^\prime},\omega)$ using the T-matrix formalism,

\begin{equation}
    G(\mathbf{R},\mathbf{R'},\omega) = G^0(\mathbf{R} -\mathbf{R'},\omega) + G^0(\mathbf{R},\omega)T(\omega) G^0( -\mathbf{R'},\omega)
    \label{Eq:GreensFunction_Tmatrix}
\end{equation}

\noindent where $T(\omega)$ describes the scattering from the point-like defect 
\begin{equation}
    T(\omega) = \hat{V}[\hat{1} - \hat{V}G^0(\mathbf{R=0},\omega)]^{-1}.
    \label{Eq:Tmatrix}
\end{equation} 
Here, we use $\hat{V}=100$~meV. 
$G^0( \mathbf{R},\omega)$ is the Fourier transform of the non-interacting Greens function $G^0( \mathbf{k},\omega)$ obtained from 

\begin{equation}
    G^0(\mathbf{k},\omega) = [(\omega + i\Gamma)\hat{1} - H(\mathbf{k})]^{-1}
    \label{Eq:GreensFunction_kspace}
\end{equation}

Here, we calculate $G^0(\mathbf{R})$ via Fourier transform of $G^0(\mathbf{k},\omega)$ on a discretised k-grid of 512 k-points in each dimension, and fixing the energy broadening parameter $\Gamma=1$~meV.

From the lattice Green's function, we obtain the continuum Green's function $G(\mathbf{r},\mathbf{r'},\omega)$ via a continuum transformation,
\begin{equation}
    G(\mathbf{r},\mathbf{r'},\omega) = \sum_{R} \sum_{R'} G(\mathbf{R},\mathbf{R'},\omega) W(\mathbf{R-r})W(\mathbf{R'-r'}).
    \label{Eq:Continuum_Transformation}
\end{equation}

Here $\mathbf{R}$ is the discrete lattice vector and $W(\mathbf{R-r})$ is the localized Wannier function required for the transformation to the continuum. 

We then obtain the continuum local density of states (cLDOS) from the continuum Green's function through

\begin{equation}
    \rho(\mathbf{r},\omega) = -\frac{1}{\pi}\mathrm{Im} G(\mathbf{r},\mathbf{r},\omega)
    \label{Eq:LDOS_supp}
\end{equation}

where the $\mathbf{r}$ is the continuous real space vector and $\omega$ the energy.

For the nearest neighbour cubic model discussed here, we use an isotropic s-wave Gaussian,

\begin{equation}
    W(\mathbf{r}) = e^{-\mathbf{r}^2/2C^2 \sigma^2},
    \label{Eq:WannierFunction}
\end{equation}

with $C=\sqrt{2\ln{100}}$ as the Wannier function. A Gaussian of this form ensures that the correct radial decay for the atomic wavefunctions at surfaces is captured, whilst allowing us to tune the overall radius of the Gaussian by using the parameter $\sigma$, where $W(\mathbf{r}=\sigma) = 0.01$. Here we choose a value of $\sigma = 1.8$, in units of the lattice constant, to ensure sufficient overlap between nearest neighbour atoms, whilst ensuring next-nearest neighbour overlap can be neglected. The choice of this parameter does not affect the qualitative behaviour of the QPI vectors, only their relative intensities. To illustrate this, we plot in Fig.~\ref{Fig:Supp_RadiusVscQPI} the cQPI for bulk and surface defects for a variety of Wannier radii $\sigma$.

\begin{figure*}
    \begin{center}
    \includegraphics[width=0.8\linewidth]{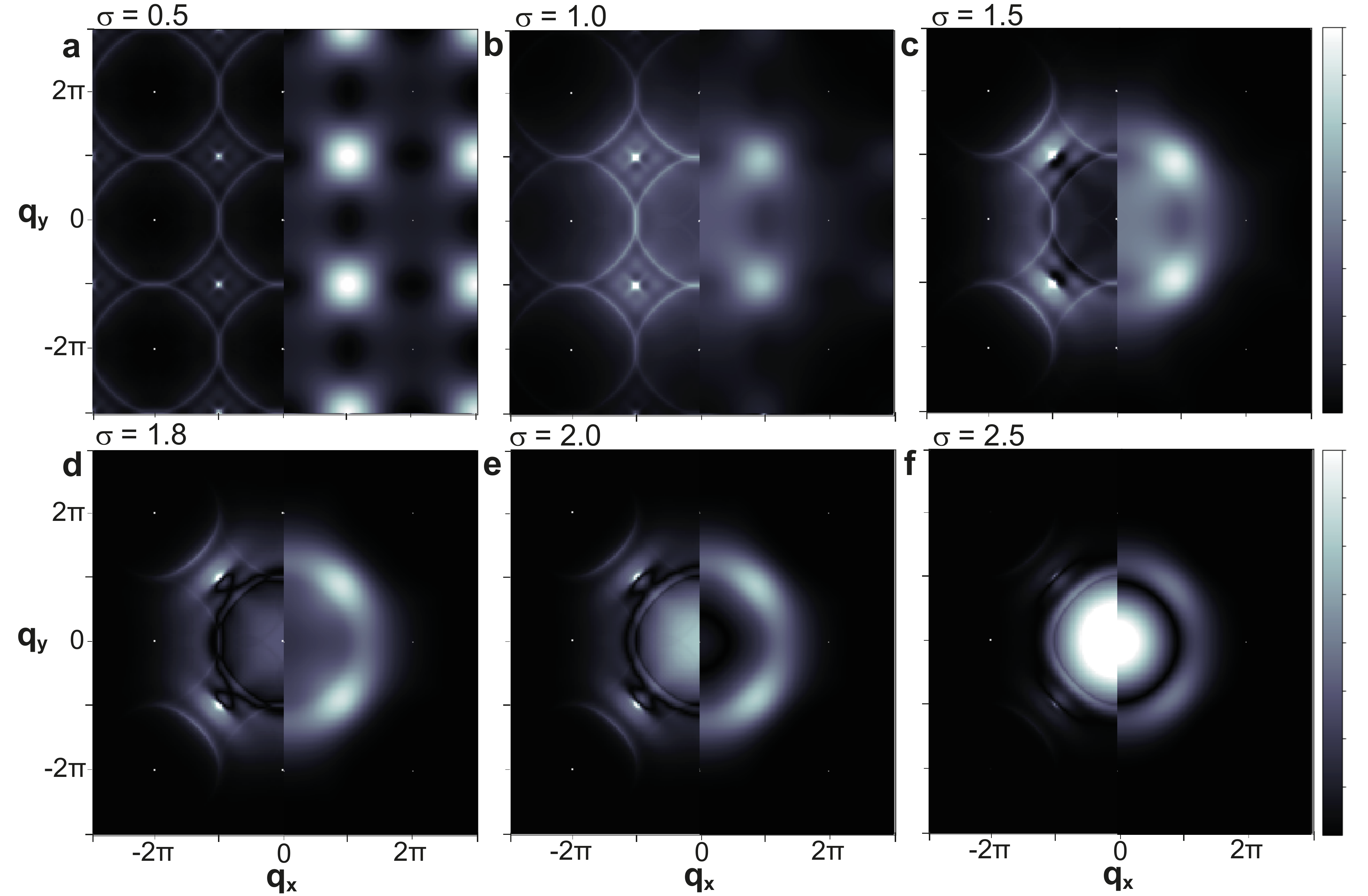}
    \end{center}
    \caption{The effect of the Wannier radius ($\sigma$) on the simulated cQPI patterns for the nearest neighbour cubic model, in units of the lattice constant.  a) $\sigma=0.5$, b) $\sigma=1.0$, c) $\sigma=1.5$, d) $\sigma=1.8$, e) $\sigma=2.0$, f) $\sigma=2.5$. The left hand side of each image is calculated for a bulk defect, whilst the right hand side is calculated for a surface defect.  For small $\sigma$ ($\sigma < 0.5$) the cQPI pattern approaches the discretised case obtained directly from the imaginary part of Eq. \eqref{Eq:GreensFunction_Tmatrix}. Whereas for large $\sigma$ ($\sigma < 2.0$), the intensity is localised and dominates around $q=0$. Note a fixed height of $r_z=0$ was used here, rather than $r_z=1.5$ unit cells in the main text, to enable sufficient overlap between the wavefunctions of the atoms and the measurement region.}
    \label{Fig:Supp_RadiusVscQPI}
\end{figure*}

\begin{figure*}
    \begin{center}
    \includegraphics[width=0.9\linewidth]{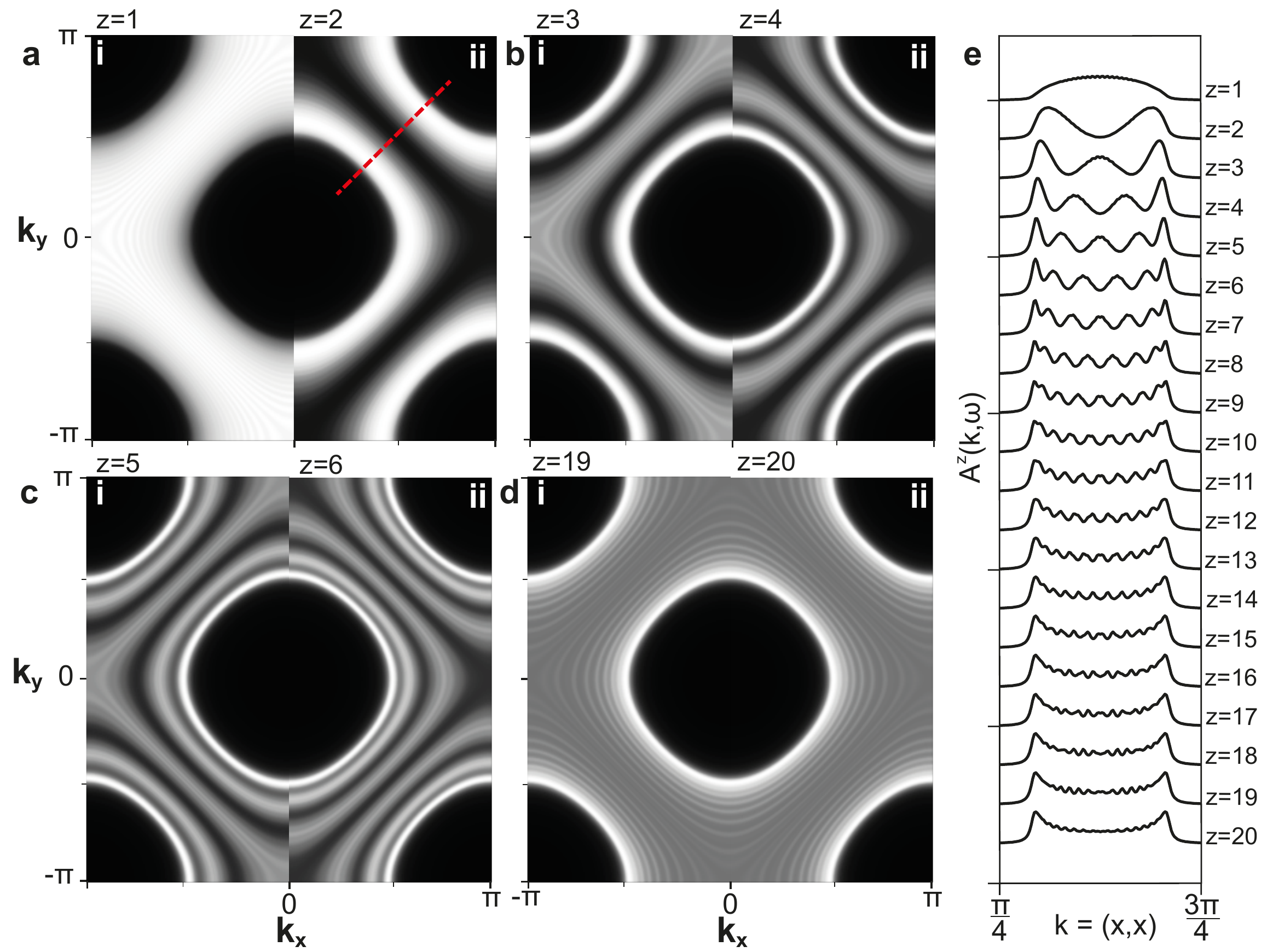}
    \end{center}
    \caption{Depth dependence of the spectral function (Eq. \eqref{Eq:PartialSpectralFunction}) on the nearest neighbour cubic model for a 40 layer slab system. ai),aii) The surface (z=1) and subsurface (z=2) partial spectral function. bi),bii) The partial spectral function of the z=3 and z=4 layers, ci),cii) The partial spectral function of the z=5 and z=6 layers, di),dii) The partial spectral function of the z=19 and z=20 layers, which approximates the bulk of the material. e) Surface to bulk progression of the spectral intensity along the $\mathbf{k}$-path shown as the red dashed line in (aii). For the bulk like layers, the spectral intensity is equivalent to the summation of the full three dimensional electronic structure over all $k_z$. For small z, certain momentum have enhanced spectral weight, this correspond to different regions of $k_z$ from the full three-dimensional electronic structure. This layer dependence can then be used to understand $k_z$ from two dimensional surface probes.}
    \label{Fig:DepthSpectral}
\end{figure*}

\section{Depth dependence of the spectral function}
In Fig.~\ref{Fig:DepthSpectral}, we show the partial spectral function of the individual layers for a 40 layer slab Hamiltonian of the simple cubic isotropic nearest neighbour hopping model, as defined by Eq. (4) of the main text

\begin{equation}
H(\mathbf{k_\parallel}) = 
\begin{pmatrix}
   H^{0}(\mathbf{k_\parallel}) & H^{1}(\mathbf{k_\parallel}) & H^{2}(\mathbf{k_\parallel}) &  ... \\
   H^{1}(\mathbf{k_\parallel}) & H^{0}(\mathbf{k_\parallel}) & H^{1}(\mathbf{k_\parallel}) &   ... \\
     H^{2}(\mathbf{k_\parallel}) & H^{1}(\mathbf{k_\parallel}) &H^{0}(\mathbf{k_\parallel}) & ... \\
     ... & ... & ... & ... \\
    \end{pmatrix}.
\label{Eq:Ham_Slab}
\end{equation}

We then calculate the corresponding Greens function using Eq.~\eqref{Eq:GreensFunction_kspace} and look at the imaginary part of the $z$-th diagonal element to obtain the spectral function at different depths

\begin{equation}
A^z(\mathbf{k_\parallel,\omega}) = -\frac{1}{\pi} \mathrm{Im} G^{zz}(\mathbf{k_\parallel,\omega}).
\label{Eq:PartialSpectralFunction}
\end{equation}

By comparing this with the full three dimensional electronic structure shown in Fig.~1(g) of the main text, it can be seen that the bulk ($z=20$) corresponds to a summation of all $k_z$ states, with the greatest intensity arising for states where the Fermi velocity in $k_z$ is smallest. However for the surface, although the intensity spans the entire range of $k_\parallel$ measured by the bulk, the intensity does not correspond to a $k_z$ summation but shows a pronounced oscillatory behaviour (see fig.~\ref{Fig:DepthSpectral}e). The number of nodes in the spectral function increases linearly as $z$ increases. 

\section{PbS tight binding model}
DFT calculations for galena (PbS), which has a rocksalt structure (lattice constant $a=5.9315\mathrm{\AA}$), were performed using Quantum Espresso \cite{Giannozzi_2017}. The calculations have been performed using the PBE exchange correlation functional. Kinetic energy cut-offs of $40 \mathrm{Ry}$, and $320\mathrm{Ry}$ were used for the wavefunction and charge density, respectively. We used a $6\times 6\times 6$ Monkhurst-Pack $k$-grid. This DFT bandstructure was then projected into an orthogonal tight-binding model using Wannier90 \cite{Pizzi2020}. This Hamiltonian contained the $p_x$, $p_y$ and $p_z$ orbitals on each of the four Pb and S sites generating a 24-orbital Hamiltonian. A $4\times 4\times 4$ Monkhurst-Pack grid was used for the Wannierisation. The chosen energy windows, in units of $\mathrm{eV}$, for disentanglement were $[-10.9705,9.2295]$ and frozen window $[-5.4705,3.2295]$ with respect to the Fermi energy. The original input files can be found in ref.~\cite{3dqpidata}. This tight-binding model was then converted into a 16 layer slab following the method described of the main text. 

For the cQPI calculations, we approximated the Wannier functions using Slater-Type orbitals $W_i = \sqrt{\frac{3}{4 \pi}} \frac{i}{r} e^{\frac{-|\mathbf{r}|}{\sigma}}$, where $i = x,y$ or $z$ defines the specific $p$-orbital symmetry, and set $\sigma = 0.25$ unit cells to reproduce the q-intensity decay observed in experiment. 
\end{widetext}


\begin{thebibliography}{32}%
	\makeatletter
	\providecommand \@ifxundefined [1]{%
		\@ifx{#1\undefined}
	}%
	\providecommand \@ifnum [1]{%
		\ifnum #1\expandafter \@firstoftwo
		\else \expandafter \@secondoftwo
		\fi
	}%
	\providecommand \@ifx [1]{%
		\ifx #1\expandafter \@firstoftwo
		\else \expandafter \@secondoftwo
		\fi
	}%
	\providecommand \natexlab [1]{#1}%
	\providecommand \enquote  [1]{``#1''}%
	\providecommand \bibnamefont  [1]{#1}%
	\providecommand \bibfnamefont [1]{#1}%
	\providecommand \citenamefont [1]{#1}%
	\providecommand \href@noop [0]{\@secondoftwo}%
	\providecommand \href [0]{\begingroup \@sanitize@url \@href}%
	\providecommand \@href[1]{\@@startlink{#1}\@@href}%
	\providecommand \@@href[1]{\endgroup#1\@@endlink}%
	\providecommand \@sanitize@url [0]{\catcode `\\12\catcode `\$12\catcode
		`\&12\catcode `\#12\catcode `\^12\catcode `\_12\catcode `\%12\relax}%
	\providecommand \@@startlink[1]{}%
	\providecommand \@@endlink[0]{}%
	\providecommand \url  [0]{\begingroup\@sanitize@url \@url }%
	\providecommand \@url [1]{\endgroup\@href {#1}{\urlprefix }}%
	\providecommand \urlprefix  [0]{URL }%
	\providecommand \Eprint [0]{\href }%
	\providecommand \doibase [0]{http://dx.doi.org/}%
	\providecommand \selectlanguage [0]{\@gobble}%
	\providecommand \bibinfo  [0]{\@secondoftwo}%
	\providecommand \bibfield  [0]{\@secondoftwo}%
	\providecommand \translation [1]{[#1]}%
	\providecommand \BibitemOpen [0]{}%
	\providecommand \bibitemStop [0]{}%
	\providecommand \bibitemNoStop [0]{.\EOS\space}%
	\providecommand \EOS [0]{\spacefactor3000\relax}%
	\providecommand \BibitemShut  [1]{\csname bibitem#1\endcsname}%
	\let\auto@bib@innerbib\@empty
	\bibitem [{\citenamefont {Crommie}\ \emph {et~al.}(1993)\citenamefont
		{Crommie}, \citenamefont {Lutz},\ and\ \citenamefont {Eigler}}]{Crommie1993}%
	\BibitemOpen
	\bibfield  {author} {\bibinfo {author} {\bibfnamefont {M.~F.}\ \bibnamefont
			{Crommie}}, \bibinfo {author} {\bibfnamefont {C.~P.}\ \bibnamefont {Lutz}}, \
		and\ \bibinfo {author} {\bibfnamefont {D.~M.}\ \bibnamefont {Eigler}},\
	}\bibfield  {title} {\enquote {\bibinfo {title} {{Imaging standing waves in a
					two-dimensional electron gas}},}\ }\href {\doibase 10.1038/363524a0}
	{\bibfield  {journal} {\bibinfo  {journal} {Nature}\ }\textbf {\bibinfo
			{volume} {363}},\ \bibinfo {pages} {524--527} (\bibinfo {year}
		{1993})}\BibitemShut {NoStop}%
	\bibitem [{\citenamefont {Hasegawa}\ and\ \citenamefont
		{Avouris}(1993)}]{hasegawa_direct_1993}%
	\BibitemOpen
	\bibfield  {author} {\bibinfo {author} {\bibfnamefont {Y.}~\bibnamefont
			{Hasegawa}}\ and\ \bibinfo {author} {\bibfnamefont {Ph.}\ \bibnamefont
			{Avouris}},\ }\bibfield  {title} {\enquote {\bibinfo {title} {{Direct
					observation of standing wave formation at surface steps using scanning
					tunneling spectroscopy}},}\ }\href {\doibase 10.1103/PhysRevLett.71.1071}
	{\bibfield  {journal} {\bibinfo  {journal} {Phys. Rev. Lett.}\ }\textbf
		{\bibinfo {volume} {71}},\ \bibinfo {pages} {1071--1074} (\bibinfo {year}
		{1993})}\BibitemShut {NoStop}%
	\bibitem [{\citenamefont {Ast}\ \emph {et~al.}(2016)\citenamefont {Ast},
		\citenamefont {J{\"a}ck}, \citenamefont {Senkpiel}, \citenamefont {Eltschka},
		\citenamefont {Etzkorn}, \citenamefont {Ankerhold},\ and\ \citenamefont
		{Kern}}]{ast_sensing_2016}%
	\BibitemOpen
	\bibfield  {author} {\bibinfo {author} {\bibfnamefont {C.~R.}\ \bibnamefont
			{Ast}}, \bibinfo {author} {\bibfnamefont {B.}~\bibnamefont {J{\"a}ck}},
		\bibinfo {author} {\bibfnamefont {J.}~\bibnamefont {Senkpiel}}, \bibinfo
		{author} {\bibfnamefont {M.}~\bibnamefont {Eltschka}}, \bibinfo {author}
		{\bibfnamefont {M.}~\bibnamefont {Etzkorn}}, \bibinfo {author} {\bibfnamefont
			{J.}~\bibnamefont {Ankerhold}}, \ and\ \bibinfo {author} {\bibfnamefont
			{K.}~\bibnamefont {Kern}},\ }\bibfield  {title} {\enquote {\bibinfo {title}
			{{Sensing the quantum limit in scanning tunnelling spectroscopy}},}\ }\href
	{\doibase 10.1038/ncomms13009} {\bibfield  {journal} {\bibinfo  {journal}
			{Nat. Comm.}\ }\textbf {\bibinfo {volume} {7}},\ \bibinfo {pages} {13009}
		(\bibinfo {year} {2016})}\BibitemShut {NoStop}%
	\bibitem [{\citenamefont {Petersen}\ \emph {et~al.}(1998)\citenamefont
		{Petersen}, \citenamefont {Laitenberger}, \citenamefont {L{\ae}gsgaard},\
		and\ \citenamefont {Besenbacher}}]{Petersen1998}%
	\BibitemOpen
	\bibfield  {author} {\bibinfo {author} {\bibfnamefont {L.}~\bibnamefont
			{Petersen}}, \bibinfo {author} {\bibfnamefont {P.}~\bibnamefont
			{Laitenberger}}, \bibinfo {author} {\bibfnamefont {E.}~\bibnamefont
			{L{\ae}gsgaard}}, \ and\ \bibinfo {author} {\bibfnamefont {F.}~\bibnamefont
			{Besenbacher}},\ }\bibfield  {title} {\enquote {\bibinfo {title} {{Screening
					waves from steps and defects on Cu(111) and Au(111) imaged with
					STM:Contribution from bulk electrons}},}\ }\href {\doibase
		10.1103/physrevb.58.7361} {\bibfield  {journal} {\bibinfo  {journal} {Phys.
				Rev. B}\ }\textbf {\bibinfo {volume} {58}},\ \bibinfo {pages} {7361--7366}
		(\bibinfo {year} {1998})}\BibitemShut {NoStop}%
	\bibitem [{\citenamefont {Simon}\ \emph {et~al.}(2011)\citenamefont {Simon},
		\citenamefont {Bena}, \citenamefont {Vonau}, \citenamefont {Cranney},\ and\
		\citenamefont {Aubel}}]{Simon_2011}%
	\BibitemOpen
	\bibfield  {author} {\bibinfo {author} {\bibfnamefont {L}~\bibnamefont
			{Simon}}, \bibinfo {author} {\bibfnamefont {C}~\bibnamefont {Bena}}, \bibinfo
		{author} {\bibfnamefont {F}~\bibnamefont {Vonau}}, \bibinfo {author}
		{\bibfnamefont {M}~\bibnamefont {Cranney}}, \ and\ \bibinfo {author}
		{\bibfnamefont {D}~\bibnamefont {Aubel}},\ }\bibfield  {title} {\enquote
		{\bibinfo {title} {Fourier-transform scanning tunnelling spectroscopy: the
				possibility to obtain constant-energy maps and band dispersion using a local
				measurement},}\ }\href {\doibase 10.1088/0022-3727/44/46/464010} {\bibfield
		{journal} {\bibinfo  {journal} {Journal of Physics D: Applied Physics}\
		}\textbf {\bibinfo {volume} {44}},\ \bibinfo {pages} {464010} (\bibinfo
		{year} {2011})}\BibitemShut {NoStop}%
	\bibitem [{\citenamefont {Hoffman}(2002)}]{Hoffman2002}%
	\BibitemOpen
	\bibfield  {author} {\bibinfo {author} {\bibfnamefont {J.~E.}\ \bibnamefont
			{Hoffman}},\ }\bibfield  {title} {\enquote {\bibinfo {title} {Imaging
				quasiparticle interference in {Bi}$_2${Sr}$_2${CaCu}$_2${O}$_{8+\delta}$},}\
	}\href {\doibase 10.1126/science.1072640} {\bibfield  {journal} {\bibinfo
			{journal} {Science}\ }\textbf {\bibinfo {volume} {297}},\ \bibinfo {pages}
		{1148--1151} (\bibinfo {year} {2002})}\BibitemShut {NoStop}%
	\bibitem [{\citenamefont {Kohsaka}\ \emph {et~al.}(2008)\citenamefont
		{Kohsaka}, \citenamefont {Taylor}, \citenamefont {Wahl}, \citenamefont
		{Schmidt}, \citenamefont {Lee}, \citenamefont {Fujita}, \citenamefont
		{Alldredge}, \citenamefont {McElroy}, \citenamefont {Lee}, \citenamefont
		{Eisaki}, \citenamefont {Uchida}, \citenamefont {Lee},\ and\ \citenamefont
		{Davis}}]{kohsaka_how_2008}%
	\BibitemOpen
	\bibfield  {author} {\bibinfo {author} {\bibfnamefont {Y.}~\bibnamefont
			{Kohsaka}}, \bibinfo {author} {\bibfnamefont {C.}~\bibnamefont {Taylor}},
		\bibinfo {author} {\bibfnamefont {P.}~\bibnamefont {Wahl}}, \bibinfo {author}
		{\bibfnamefont {A.}~\bibnamefont {Schmidt}}, \bibinfo {author} {\bibfnamefont
			{Jhinhwan}\ \bibnamefont {Lee}}, \bibinfo {author} {\bibfnamefont
			{K.}~\bibnamefont {Fujita}}, \bibinfo {author} {\bibfnamefont {J.~W.}\
			\bibnamefont {Alldredge}}, \bibinfo {author} {\bibfnamefont {K.}~\bibnamefont
			{McElroy}}, \bibinfo {author} {\bibfnamefont {Jinho}\ \bibnamefont {Lee}},
		\bibinfo {author} {\bibfnamefont {H.}~\bibnamefont {Eisaki}}, \bibinfo
		{author} {\bibfnamefont {S.}~\bibnamefont {Uchida}}, \bibinfo {author}
		{\bibfnamefont {D.-H.}\ \bibnamefont {Lee}}, \ and\ \bibinfo {author}
		{\bibfnamefont {J.~C.}\ \bibnamefont {Davis}},\ }\bibfield  {title} {\enquote
		{\bibinfo {title} {{How Cooper pairs vanish approaching the {Mott} insulator
					in Bi$_2$Sr$_2$CaCu$_2$O$_{8+\delta}$}},}\ }\href {\doibase
		10.1038/nature07243} {\bibfield  {journal} {\bibinfo  {journal} {Nature}\
		}\textbf {\bibinfo {volume} {454}},\ \bibinfo {pages} {1072--1078} (\bibinfo
		{year} {2008})}\BibitemShut {NoStop}%
	\bibitem [{\citenamefont {He}\ \emph {et~al.}(2014)\citenamefont {He},
		\citenamefont {Yin}, \citenamefont {Zech}, \citenamefont {Soumyanarayanan},
		\citenamefont {Yee}, \citenamefont {Williams}, \citenamefont {Boyer},
		\citenamefont {Chatterjee}, \citenamefont {Wise}, \citenamefont {Zeljkovic},
		\citenamefont {Kondo}, \citenamefont {Takeuchi}, \citenamefont {Ikuta},
		\citenamefont {Mistark}, \citenamefont {Markiewicz}, \citenamefont {Bansil},
		\citenamefont {Sachdev}, \citenamefont {Hudson},\ and\ \citenamefont
		{Hoffman}}]{He2014_Cuprate}%
	\BibitemOpen
	\bibfield  {author} {\bibinfo {author} {\bibfnamefont {Y.}~\bibnamefont
			{He}}, \bibinfo {author} {\bibfnamefont {Y.}~\bibnamefont {Yin}}, \bibinfo
		{author} {\bibfnamefont {M.}~\bibnamefont {Zech}}, \bibinfo {author}
		{\bibfnamefont {A.}~\bibnamefont {Soumyanarayanan}}, \bibinfo {author}
		{\bibfnamefont {M.~M.}\ \bibnamefont {Yee}}, \bibinfo {author} {\bibfnamefont
			{T.}~\bibnamefont {Williams}}, \bibinfo {author} {\bibfnamefont {M.~C.}\
			\bibnamefont {Boyer}}, \bibinfo {author} {\bibfnamefont {K.}~\bibnamefont
			{Chatterjee}}, \bibinfo {author} {\bibfnamefont {W.~D.}\ \bibnamefont
			{Wise}}, \bibinfo {author} {\bibfnamefont {I.}~\bibnamefont {Zeljkovic}},
		\bibinfo {author} {\bibfnamefont {T.}~\bibnamefont {Kondo}}, \bibinfo
		{author} {\bibfnamefont {T.}~\bibnamefont {Takeuchi}}, \bibinfo {author}
		{\bibfnamefont {H.}~\bibnamefont {Ikuta}}, \bibinfo {author} {\bibfnamefont
			{P.}~\bibnamefont {Mistark}}, \bibinfo {author} {\bibfnamefont {R.~S.}\
			\bibnamefont {Markiewicz}}, \bibinfo {author} {\bibfnamefont
			{A.}~\bibnamefont {Bansil}}, \bibinfo {author} {\bibfnamefont
			{S.}~\bibnamefont {Sachdev}}, \bibinfo {author} {\bibfnamefont {E.~W.}\
			\bibnamefont {Hudson}}, \ and\ \bibinfo {author} {\bibfnamefont {J.~E.}\
			\bibnamefont {Hoffman}},\ }\bibfield  {title} {\enquote {\bibinfo {title}
			{{Fermi Surface and Pseudogap Evolution in a Cuprate Superconductor}},}\
	}\href {\doibase 10.1126/science.1248221} {\bibfield  {journal} {\bibinfo
			{journal} {Science}\ }\textbf {\bibinfo {volume} {344}},\ \bibinfo {pages}
		{608--611} (\bibinfo {year} {2014})}\BibitemShut {NoStop}%
	\bibitem [{\citenamefont {Wang}\ \emph {et~al.}(2017)\citenamefont {Wang},
		\citenamefont {Walkup}, \citenamefont {Derry}, \citenamefont {Scaffidi},
		\citenamefont {Rak}, \citenamefont {Vig}, \citenamefont {Kogar},
		\citenamefont {Zeljkovic}, \citenamefont {Husain}, \citenamefont {Santos},
		\citenamefont {Wang}, \citenamefont {Damascelli}, \citenamefont {Maeno},
		\citenamefont {Abbamonte}, \citenamefont {Fradkin},\ and\ \citenamefont
		{Madhavan}}]{wang_quasiparticle_2017}%
	\BibitemOpen
	\bibfield  {author} {\bibinfo {author} {\bibfnamefont {Z.}~\bibnamefont
			{Wang}}, \bibinfo {author} {\bibfnamefont {D.}~\bibnamefont {Walkup}},
		\bibinfo {author} {\bibfnamefont {P.}~\bibnamefont {Derry}}, \bibinfo
		{author} {\bibfnamefont {T.}~\bibnamefont {Scaffidi}}, \bibinfo {author}
		{\bibfnamefont {M.}~\bibnamefont {Rak}}, \bibinfo {author} {\bibfnamefont
			{S.}~\bibnamefont {Vig}}, \bibinfo {author} {\bibfnamefont {A.}~\bibnamefont
			{Kogar}}, \bibinfo {author} {\bibfnamefont {I.}~\bibnamefont {Zeljkovic}},
		\bibinfo {author} {\bibfnamefont {A.}~\bibnamefont {Husain}}, \bibinfo
		{author} {\bibfnamefont {L.~H.}\ \bibnamefont {Santos}}, \bibinfo {author}
		{\bibfnamefont {Y.}~\bibnamefont {Wang}}, \bibinfo {author} {\bibfnamefont
			{A.}~\bibnamefont {Damascelli}}, \bibinfo {author} {\bibfnamefont
			{Y.}~\bibnamefont {Maeno}}, \bibinfo {author} {\bibfnamefont
			{P.}~\bibnamefont {Abbamonte}}, \bibinfo {author} {\bibfnamefont
			{E.}~\bibnamefont {Fradkin}}, \ and\ \bibinfo {author} {\bibfnamefont
			{V.}~\bibnamefont {Madhavan}},\ }\bibfield  {title} {\enquote {\bibinfo
			{title} {{Quasiparticle interference and strong electron–mode coupling in
					the quasi-one-dimensional bands of
					{Sr}$_{\textrm{2}}${RuO}$_{\textrm{4}}$}},}\ }\href {\doibase
		10.1038/nphys4107} {\bibfield  {journal} {\bibinfo  {journal} {Nature
				Physics}\ }\textbf {\bibinfo {volume} {13}},\ \bibinfo {pages} {799--805}
		(\bibinfo {year} {2017})}\BibitemShut {NoStop}%
	\bibitem [{\citenamefont {Kreisel}\ \emph {et~al.}(2021)\citenamefont
		{Kreisel}, \citenamefont {Marques}, \citenamefont {Rhodes}, \citenamefont
		{Kong}, \citenamefont {Berlijn}, \citenamefont {Fittipaldi}, \citenamefont
		{Granata}, \citenamefont {Vecchione}, \citenamefont {Wahl},\ and\
		\citenamefont {Hirschfeld}}]{Kreisel2021}%
	\BibitemOpen
	\bibfield  {author} {\bibinfo {author} {\bibfnamefont {A.}~\bibnamefont
			{Kreisel}}, \bibinfo {author} {\bibfnamefont {C.~A.}\ \bibnamefont
			{Marques}}, \bibinfo {author} {\bibfnamefont {L.~C.}\ \bibnamefont {Rhodes}},
		\bibinfo {author} {\bibfnamefont {X.}~\bibnamefont {Kong}}, \bibinfo {author}
		{\bibfnamefont {T.}~\bibnamefont {Berlijn}}, \bibinfo {author} {\bibfnamefont
			{R.}~\bibnamefont {Fittipaldi}}, \bibinfo {author} {\bibfnamefont
			{V.}~\bibnamefont {Granata}}, \bibinfo {author} {\bibfnamefont
			{A.}~\bibnamefont {Vecchione}}, \bibinfo {author} {\bibfnamefont
			{P.}~\bibnamefont {Wahl}}, \ and\ \bibinfo {author} {\bibfnamefont {P.~J.}\
			\bibnamefont {Hirschfeld}},\ }\bibfield  {title} {\enquote {\bibinfo {title}
			{{Quasi-particle interference of the van Hove singularity in
					Sr$_2$RuO$_4$}},}\ }\href {\doibase 10.1038/s41535-021-00401-x} {\bibfield
		{journal} {\bibinfo  {journal} {npj Quantum Materials}\ }\textbf {\bibinfo
			{volume} {6}},\ \bibinfo {pages} {100} (\bibinfo {year} {2021})}\BibitemShut
	{NoStop}%
	\bibitem [{\citenamefont {Allan}\ \emph {et~al.}(2012)\citenamefont {Allan},
		\citenamefont {Rost}, \citenamefont {Mackenzie}, \citenamefont {Xie},
		\citenamefont {Davis}, \citenamefont {Kihou}, \citenamefont {Lee},
		\citenamefont {Iyo}, \citenamefont {Eisaki},\ and\ \citenamefont
		{Chuang}}]{allan_anisotropic_2012}%
	\BibitemOpen
	\bibfield  {author} {\bibinfo {author} {\bibfnamefont {M.~P.}\ \bibnamefont
			{Allan}}, \bibinfo {author} {\bibfnamefont {A.~W.}\ \bibnamefont {Rost}},
		\bibinfo {author} {\bibfnamefont {A.~P.}\ \bibnamefont {Mackenzie}}, \bibinfo
		{author} {\bibfnamefont {Y.}~\bibnamefont {Xie}}, \bibinfo {author}
		{\bibfnamefont {J.~C.}\ \bibnamefont {Davis}}, \bibinfo {author}
		{\bibfnamefont {K.}~\bibnamefont {Kihou}}, \bibinfo {author} {\bibfnamefont
			{C.~H.}\ \bibnamefont {Lee}}, \bibinfo {author} {\bibfnamefont
			{A.}~\bibnamefont {Iyo}}, \bibinfo {author} {\bibfnamefont {H.}~\bibnamefont
			{Eisaki}}, \ and\ \bibinfo {author} {\bibfnamefont {T.-M.}\ \bibnamefont
			{Chuang}},\ }\bibfield  {title} {\enquote {\bibinfo {title} {Anisotropic
				{Energy} {Gaps} of {Iron}-{Based} {Superconductivity} from {Intraband}
				{Quasiparticle} {Interference} in {LiFeAs}},}\ }\href {\doibase
		10.1126/science.1218726} {\bibfield  {journal} {\bibinfo  {journal}
			{Science}\ }\textbf {\bibinfo {volume} {336}},\ \bibinfo {pages} {563--567}
		(\bibinfo {year} {2012})}\BibitemShut {NoStop}%
	\bibitem [{\citenamefont {Allan}\ \emph {et~al.}(2014)\citenamefont {Allan},
		\citenamefont {Lee}, \citenamefont {Rost}, \citenamefont {Fischer},
		\citenamefont {Massee}, \citenamefont {Kihou}, \citenamefont {Lee},
		\citenamefont {Iyo}, \citenamefont {Eisaki}, \citenamefont {Chuang},
		\citenamefont {Davis},\ and\ \citenamefont {Kim}}]{allan_identifying_2014}%
	\BibitemOpen
	\bibfield  {author} {\bibinfo {author} {\bibfnamefont {M.~P.}\ \bibnamefont
			{Allan}}, \bibinfo {author} {\bibfnamefont {K.}~\bibnamefont {Lee}}, \bibinfo
		{author} {\bibfnamefont {A.~W.}\ \bibnamefont {Rost}}, \bibinfo {author}
		{\bibfnamefont {M.~H.}\ \bibnamefont {Fischer}}, \bibinfo {author}
		{\bibfnamefont {F.}~\bibnamefont {Massee}}, \bibinfo {author} {\bibfnamefont
			{K.}~\bibnamefont {Kihou}}, \bibinfo {author} {\bibfnamefont {C-H.}\
			\bibnamefont {Lee}}, \bibinfo {author} {\bibfnamefont {A.}~\bibnamefont
			{Iyo}}, \bibinfo {author} {\bibfnamefont {H.}~\bibnamefont {Eisaki}},
		\bibinfo {author} {\bibfnamefont {T-M.}\ \bibnamefont {Chuang}}, \bibinfo
		{author} {\bibfnamefont {J.~C.}\ \bibnamefont {Davis}}, \ and\ \bibinfo
		{author} {\bibfnamefont {E-A.}\ \bibnamefont {Kim}},\ }\bibfield  {title}
	{\enquote {\bibinfo {title} {Identifying the 'fingerprint' of
				antiferromagnetic spin fluctuations in iron pnictide superconductors},}\
	}\href {\doibase 10.1038/nphys3187} {\bibfield  {journal} {\bibinfo
			{journal} {Nature Physics}\ }\textbf {\bibinfo {volume} {11}},\ \bibinfo
		{pages} {177--182} (\bibinfo {year} {2014})}\BibitemShut {NoStop}%
	\bibitem [{\citenamefont {Sprau}\ \emph {et~al.}(2017)\citenamefont {Sprau},
		\citenamefont {Kostin}, \citenamefont {Kreisel}, \citenamefont
		{B{\"{o}}hmer}, \citenamefont {Taufour}, \citenamefont {Canfield},
		\citenamefont {Mukherjee}, \citenamefont {Hirschfeld}, \citenamefont
		{Andersen},\ and\ \citenamefont {S{\'{e}}amus~Davis}}]{Sprau2017}%
	\BibitemOpen
	\bibfield  {author} {\bibinfo {author} {\bibfnamefont {P.~O.}\ \bibnamefont
			{Sprau}}, \bibinfo {author} {\bibfnamefont {A.}~\bibnamefont {Kostin}},
		\bibinfo {author} {\bibfnamefont {A.}~\bibnamefont {Kreisel}}, \bibinfo
		{author} {\bibfnamefont {A.~E.}\ \bibnamefont {B{\"{o}}hmer}}, \bibinfo
		{author} {\bibfnamefont {V.}~\bibnamefont {Taufour}}, \bibinfo {author}
		{\bibfnamefont {P.~C.}\ \bibnamefont {Canfield}}, \bibinfo {author}
		{\bibfnamefont {S.}~\bibnamefont {Mukherjee}}, \bibinfo {author}
		{\bibfnamefont {P.~J.}\ \bibnamefont {Hirschfeld}}, \bibinfo {author}
		{\bibfnamefont {B.~M.}\ \bibnamefont {Andersen}}, \ and\ \bibinfo {author}
		{\bibfnamefont {J.~C.}\ \bibnamefont {S{\'{e}}amus~Davis}},\ }\bibfield
	{title} {\enquote {\bibinfo {title} {{Discovery of Orbital-Selective Cooper
					Pairing in FeSe}},}\ }\href {\doibase 10.1126/science.aal1575} {\bibfield
		{journal} {\bibinfo  {journal} {Science}\ }\textbf {\bibinfo {volume}
			{357}},\ \bibinfo {pages} {75--80} (\bibinfo {year} {2017})}\BibitemShut
	{NoStop}%
	\bibitem [{\citenamefont {Schmidt}\ \emph {et~al.}(2010)\citenamefont
		{Schmidt}, \citenamefont {Hamidian}, \citenamefont {Wahl}, \citenamefont
		{Meier}, \citenamefont {Balatsky}, \citenamefont {Garrett}, \citenamefont
		{Williams}, \citenamefont {Luke},\ and\ \citenamefont
		{Davis}}]{schmidt_imaging_2010}%
	\BibitemOpen
	\bibfield  {author} {\bibinfo {author} {\bibfnamefont {A.~R.}\ \bibnamefont
			{Schmidt}}, \bibinfo {author} {\bibfnamefont {M.~H.}\ \bibnamefont
			{Hamidian}}, \bibinfo {author} {\bibfnamefont {P.}~\bibnamefont {Wahl}},
		\bibinfo {author} {\bibfnamefont {F.}~\bibnamefont {Meier}}, \bibinfo
		{author} {\bibfnamefont {A.~V.}\ \bibnamefont {Balatsky}}, \bibinfo {author}
		{\bibfnamefont {J.~D.}\ \bibnamefont {Garrett}}, \bibinfo {author}
		{\bibfnamefont {T.~J.}\ \bibnamefont {Williams}}, \bibinfo {author}
		{\bibfnamefont {G.~M.}\ \bibnamefont {Luke}}, \ and\ \bibinfo {author}
		{\bibfnamefont {J.~C.}\ \bibnamefont {Davis}},\ }\bibfield  {title} {\enquote
		{\bibinfo {title} {{Imaging the {Fano} lattice to ‘hidden order’
					transition in URu$_2$Si$_2$}},}\ }\href {\doibase 10.1038/nature09073}
	{\bibfield  {journal} {\bibinfo  {journal} {Nature}\ }\textbf {\bibinfo
			{volume} {465}},\ \bibinfo {pages} {570--576} (\bibinfo {year}
		{2010})}\BibitemShut {NoStop}%
	\bibitem [{\citenamefont {Zhou}\ \emph {et~al.}(2013)\citenamefont {Zhou},
		\citenamefont {Misra}, \citenamefont {da~Silva~Neto}, \citenamefont
		{Aynajian}, \citenamefont {Baumbach}, \citenamefont {Thompson}, \citenamefont
		{Bauer},\ and\ \citenamefont {Yazdani}}]{zhou_visualizing_2013}%
	\BibitemOpen
	\bibfield  {author} {\bibinfo {author} {\bibfnamefont {B.~B.}\ \bibnamefont
			{Zhou}}, \bibinfo {author} {\bibfnamefont {S.}~\bibnamefont {Misra}},
		\bibinfo {author} {\bibfnamefont {E.~H.}\ \bibnamefont {da~Silva~Neto}},
		\bibinfo {author} {\bibfnamefont {P.}~\bibnamefont {Aynajian}}, \bibinfo
		{author} {\bibfnamefont {R.~E.}\ \bibnamefont {Baumbach}}, \bibinfo {author}
		{\bibfnamefont {J.~D.}\ \bibnamefont {Thompson}}, \bibinfo {author}
		{\bibfnamefont {E.~D.}\ \bibnamefont {Bauer}}, \ and\ \bibinfo {author}
		{\bibfnamefont {A.}~\bibnamefont {Yazdani}},\ }\bibfield  {title} {\enquote
		{\bibinfo {title} {Visualizing nodal heavy fermion superconductivity in
				{CeCoIn$_5$}},}\ }\href {\doibase 10.1038/nphys2672} {\bibfield  {journal}
		{\bibinfo  {journal} {Nature Physics}\ }\textbf {\bibinfo {volume} {9}},\
		\bibinfo {pages} {474--479} (\bibinfo {year} {2013})}\BibitemShut {NoStop}%
	\bibitem [{\citenamefont {Akbari}\ \emph {et~al.}(2011)\citenamefont {Akbari},
		\citenamefont {Thalmeier},\ and\ \citenamefont {Eremin}}]{Akbari2014}%
	\BibitemOpen
	\bibfield  {author} {\bibinfo {author} {\bibfnamefont {A.}~\bibnamefont
			{Akbari}}, \bibinfo {author} {\bibfnamefont {P.}~\bibnamefont {Thalmeier}}, \
		and\ \bibinfo {author} {\bibfnamefont {I.}~\bibnamefont {Eremin}},\
	}\bibfield  {title} {\enquote {\bibinfo {title} {{Quasiparticle interference
					in the heavy-fermion superconductor CeCoIn$_5$}},}\ }\href {\doibase
		10.1103/PhysRevB.84.134505} {\bibfield  {journal} {\bibinfo  {journal} {Phys.
				Rev. B}\ }\textbf {\bibinfo {volume} {84}},\ \bibinfo {pages} {134505}
		(\bibinfo {year} {2011})}\BibitemShut {NoStop}%
	\bibitem [{\citenamefont {Hanaguri}\ \emph {et~al.}(2018)\citenamefont
		{Hanaguri}, \citenamefont {Iwaya}, \citenamefont {Kohsaka}, \citenamefont
		{Machida}, \citenamefont {Watashige}, \citenamefont {Kasahara}, \citenamefont
		{Shibauchi},\ and\ \citenamefont {Matsuda}}]{Hanaguri2018}%
	\BibitemOpen
	\bibfield  {author} {\bibinfo {author} {\bibfnamefont {T.}~\bibnamefont
			{Hanaguri}}, \bibinfo {author} {\bibfnamefont {K.}~\bibnamefont {Iwaya}},
		\bibinfo {author} {\bibfnamefont {Y.}~\bibnamefont {Kohsaka}}, \bibinfo
		{author} {\bibfnamefont {T.}~\bibnamefont {Machida}}, \bibinfo {author}
		{\bibfnamefont {T.}~\bibnamefont {Watashige}}, \bibinfo {author}
		{\bibfnamefont {S.}~\bibnamefont {Kasahara}}, \bibinfo {author}
		{\bibfnamefont {T.}~\bibnamefont {Shibauchi}}, \ and\ \bibinfo {author}
		{\bibfnamefont {Y.}~\bibnamefont {Matsuda}},\ }\bibfield  {title} {\enquote
		{\bibinfo {title} {{Two distinct superconducting pairing states divided by
					the nematic end point in
					${\mathrm{FeSe}}_{1\ensuremath{-}x}{\mathrm{S}}_{x}$}},}\ }\href
	{http://advances.sciencemag.org/content/4/5/eaar6419} {\bibfield  {journal}
		{\bibinfo  {journal} {Science Advances}\ }\textbf {\bibinfo {volume} {4}},\
		\bibinfo {pages} {eaar6419} (\bibinfo {year} {2018})}\BibitemShut {NoStop}%
	\bibitem [{\citenamefont {Rhodes}\ \emph {et~al.}(2019)\citenamefont {Rhodes},
		\citenamefont {Watson}, \citenamefont {Kim},\ and\ \citenamefont
		{Eschrig}}]{Rhodes2019}%
	\BibitemOpen
	\bibfield  {author} {\bibinfo {author} {\bibfnamefont {L.~C.}\ \bibnamefont
			{Rhodes}}, \bibinfo {author} {\bibfnamefont {M.~D.}\ \bibnamefont {Watson}},
		\bibinfo {author} {\bibfnamefont {T.~K.}\ \bibnamefont {Kim}}, \ and\
		\bibinfo {author} {\bibfnamefont {M.}~\bibnamefont {Eschrig}},\ }\bibfield
	{title} {\enquote {\bibinfo {title} {{$k_z$ Selective Scattering within
					Quasiparticle Interference Measurements of FeSe}},}\ }\href {\doibase
		10.1103/physrevlett.123.216404} {\bibfield  {journal} {\bibinfo  {journal}
			{Phys. Rev. Lett.}\ }\textbf {\bibinfo {volume} {123}},\ \bibinfo {pages}
		{216404} (\bibinfo {year} {2019})}\BibitemShut {NoStop}%
	\bibitem [{\citenamefont {Weismann}\ \emph {et~al.}(2009)\citenamefont
		{Weismann}, \citenamefont {Wenderoth}, \citenamefont {Lounis}, \citenamefont
		{Zahn}, \citenamefont {Quaas}, \citenamefont {Ulbrich}, \citenamefont
		{Dederichs},\ and\ \citenamefont {Bl{\"u}gel}}]{Weismann2009}%
	\BibitemOpen
	\bibfield  {author} {\bibinfo {author} {\bibfnamefont {A.}~\bibnamefont
			{Weismann}}, \bibinfo {author} {\bibfnamefont {M.}~\bibnamefont {Wenderoth}},
		\bibinfo {author} {\bibfnamefont {S.}~\bibnamefont {Lounis}}, \bibinfo
		{author} {\bibfnamefont {P.}~\bibnamefont {Zahn}}, \bibinfo {author}
		{\bibfnamefont {N.}~\bibnamefont {Quaas}}, \bibinfo {author} {\bibfnamefont
			{R.~G.}\ \bibnamefont {Ulbrich}}, \bibinfo {author} {\bibfnamefont {P.~H.}\
			\bibnamefont {Dederichs}}, \ and\ \bibinfo {author} {\bibfnamefont
			{S.}~\bibnamefont {Bl{\"u}gel}},\ }\bibfield  {title} {\enquote {\bibinfo
			{title} {Seeing the {F}ermi surface in real space by nanoscale electron
				focusing},}\ }\href {\doibase 10.1126/science.1168738} {\bibfield  {journal}
		{\bibinfo  {journal} {Science}\ }\textbf {\bibinfo {volume} {323}},\ \bibinfo
		{pages} {1190--1193} (\bibinfo {year} {2009})}\BibitemShut {NoStop}%
	\bibitem [{\citenamefont {Lounis}\ \emph {et~al.}(2011)\citenamefont {Lounis},
		\citenamefont {Zahn}, \citenamefont {Weismann}, \citenamefont {Wenderoth},
		\citenamefont {Ulbrich}, \citenamefont {Mertig}, \citenamefont {Dederichs},\
		and\ \citenamefont {Bl{\"u}gel}}]{Lounis2011}%
	\BibitemOpen
	\bibfield  {author} {\bibinfo {author} {\bibfnamefont {S.}~\bibnamefont
			{Lounis}}, \bibinfo {author} {\bibfnamefont {P.}~\bibnamefont {Zahn}},
		\bibinfo {author} {\bibfnamefont {A.}~\bibnamefont {Weismann}}, \bibinfo
		{author} {\bibfnamefont {M.}~\bibnamefont {Wenderoth}}, \bibinfo {author}
		{\bibfnamefont {R.~G.}\ \bibnamefont {Ulbrich}}, \bibinfo {author}
		{\bibfnamefont {I.}~\bibnamefont {Mertig}}, \bibinfo {author} {\bibfnamefont
			{P.~H.}\ \bibnamefont {Dederichs}}, \ and\ \bibinfo {author} {\bibfnamefont
			{S.}~\bibnamefont {Bl{\"u}gel}},\ }\bibfield  {title} {\enquote {\bibinfo
			{title} {{Theory of real space imaging of Fermi surface parts}},}\ }\href
	{\doibase 10.1103/physrevb.83.035427} {\bibfield  {journal} {\bibinfo
			{journal} {Phys. Rev. B}\ }\textbf {\bibinfo {volume} {83}},\ \bibinfo
		{pages} {035427} (\bibinfo {year} {2011})}\BibitemShut {NoStop}%
	\bibitem [{\citenamefont {Kotzott}\ \emph {et~al.}(2021)\citenamefont
		{Kotzott}, \citenamefont {Bouhassoune}, \citenamefont {Prüser},
		\citenamefont {Weismann}, \citenamefont {Lounis},\ and\ \citenamefont
		{Wenderoth}}]{Kotzott_2021}%
	\BibitemOpen
	\bibfield  {author} {\bibinfo {author} {\bibfnamefont {T.}~\bibnamefont
			{Kotzott}}, \bibinfo {author} {\bibfnamefont {M.}~\bibnamefont
			{Bouhassoune}}, \bibinfo {author} {\bibfnamefont {H.}~\bibnamefont
			{Prüser}}, \bibinfo {author} {\bibfnamefont {A.}~\bibnamefont {Weismann}},
		\bibinfo {author} {\bibfnamefont {S.}~\bibnamefont {Lounis}}, \ and\ \bibinfo
		{author} {\bibfnamefont {M.}~\bibnamefont {Wenderoth}},\ }\bibfield  {title}
	{\enquote {\bibinfo {title} {{Scanning tunneling spectroscopy of subsurface
					Ag and Ge impurities in copper}},}\ }\href {\doibase
		10.1088/1367-2630/ac3681} {\bibfield  {journal} {\bibinfo  {journal} {New
				Journal of Physics}\ }\textbf {\bibinfo {volume} {23}},\ \bibinfo {pages}
		{113044} (\bibinfo {year} {2021})}\BibitemShut {NoStop}%
	\bibitem [{\citenamefont {Marques}\ \emph {et~al.}(2021)\citenamefont
		{Marques}, \citenamefont {Bahramy}, \citenamefont {Trainer}, \citenamefont
		{Marković}, \citenamefont {Watson}, \citenamefont {Mazzola}, \citenamefont
		{Rajan}, \citenamefont {Raub}, \citenamefont {King},\ and\ \citenamefont
		{Wahl}}]{marques_tomographic_2021}%
	\BibitemOpen
	\bibfield  {author} {\bibinfo {author} {\bibfnamefont {C.~A.}\ \bibnamefont
			{Marques}}, \bibinfo {author} {\bibfnamefont {M.~S.}\ \bibnamefont
			{Bahramy}}, \bibinfo {author} {\bibfnamefont {C.}~\bibnamefont {Trainer}},
		\bibinfo {author} {\bibfnamefont {I.}~\bibnamefont {Marković}}, \bibinfo
		{author} {\bibfnamefont {M.~D.}\ \bibnamefont {Watson}}, \bibinfo {author}
		{\bibfnamefont {F.}~\bibnamefont {Mazzola}}, \bibinfo {author} {\bibfnamefont
			{A.}~\bibnamefont {Rajan}}, \bibinfo {author} {\bibfnamefont {T.~D.}\
			\bibnamefont {Raub}}, \bibinfo {author} {\bibfnamefont {P.~D.~C.}\
			\bibnamefont {King}}, \ and\ \bibinfo {author} {\bibfnamefont
			{P.}~\bibnamefont {Wahl}},\ }\bibfield  {title} {{\enquote {\bibinfo {title} {Tomographic mapping of the hidden dimension
					in quasi-particle interference},}\ }}\href {\doibase
		10.1038/s41467-021-27082-1} {\bibfield  {journal} {\bibinfo  {journal} {Nat
				Commun}\ }\textbf {\bibinfo {volume} {12}},\ \bibinfo {pages} {6739}
		(\bibinfo {year} {2021})}\BibitemShut {NoStop}%
	\bibitem [{\citenamefont {Choubey}\ \emph {et~al.}(2014)\citenamefont
		{Choubey}, \citenamefont {Berlijn}, \citenamefont {Kreisel}, \citenamefont
		{Cao},\ and\ \citenamefont {Hirschfeld}}]{choubey_visualization_2014}%
	\BibitemOpen
	\bibfield  {author} {\bibinfo {author} {\bibfnamefont {P.}~\bibnamefont
			{Choubey}}, \bibinfo {author} {\bibfnamefont {T.}~\bibnamefont {Berlijn}},
		\bibinfo {author} {\bibfnamefont {A.}~\bibnamefont {Kreisel}}, \bibinfo
		{author} {\bibfnamefont {C.}~\bibnamefont {Cao}}, \ and\ \bibinfo {author}
		{\bibfnamefont {P.~J.}\ \bibnamefont {Hirschfeld}},\ }\bibfield  {title}
	{\enquote {\bibinfo {title} {Visualization of atomic-scale phenomena in
				superconductors: {Application} to {FeSe}},}\ }\href {\doibase
		10.1103/PhysRevB.90.134520} {\bibfield  {journal} {\bibinfo  {journal} {Phys.
				Rev. B}\ }\textbf {\bibinfo {volume} {90}},\ \bibinfo {pages} {134520}
		(\bibinfo {year} {2014})}\BibitemShut {NoStop}%
	\bibitem [{\citenamefont {Kreisel}\ \emph {et~al.}(2015)\citenamefont
		{Kreisel}, \citenamefont {Choubey}, \citenamefont {Berlijn}, \citenamefont
		{Ku}, \citenamefont {Andersen},\ and\ \citenamefont
		{Hirschfeld}}]{Kreisel2015}%
	\BibitemOpen
	\bibfield  {author} {\bibinfo {author} {\bibfnamefont {A.}~\bibnamefont
			{Kreisel}}, \bibinfo {author} {\bibfnamefont {P.}~\bibnamefont {Choubey}},
		\bibinfo {author} {\bibfnamefont {T.}~\bibnamefont {Berlijn}}, \bibinfo
		{author} {\bibfnamefont {W.}~\bibnamefont {Ku}}, \bibinfo {author}
		{\bibfnamefont {B.~M.}\ \bibnamefont {Andersen}}, \ and\ \bibinfo {author}
		{\bibfnamefont {P.~J.}\ \bibnamefont {Hirschfeld}},\ }\bibfield  {title}
	{\enquote {\bibinfo {title} {{Interpretation of Scanning Tunneling
					Quasiparticle Interference and Impurity States in Cuprates}},}\ }\href
	{\doibase 10.1103/PhysRevLett.114.217002} {\bibfield  {journal} {\bibinfo
			{journal} {Phys. Rev. Lett.}\ }\textbf {\bibinfo {volume} {114}},\ \bibinfo
		{pages} {217002} (\bibinfo {year} {2015})}\BibitemShut {NoStop}%
	\bibitem [{\citenamefont {Lambert}\ \emph {et~al.}(2017)\citenamefont
		{Lambert}, \citenamefont {Akbari}, \citenamefont {Thalmeier},\ and\
		\citenamefont {Eremin}}]{Lambert2017}%
	\BibitemOpen
	\bibfield  {author} {\bibinfo {author} {\bibfnamefont {F.}~\bibnamefont
			{Lambert}}, \bibinfo {author} {\bibfnamefont {A.}~\bibnamefont {Akbari}},
		\bibinfo {author} {\bibfnamefont {P.}~\bibnamefont {Thalmeier}}, \ and\
		\bibinfo {author} {\bibfnamefont {I.}~\bibnamefont {Eremin}},\ }\bibfield
	{title} {\enquote {\bibinfo {title} {{Surface State Tunneling Signatures in
					the Two-Component Superconductor ${\mathrm{UPt}}_{3}$}},}\ }\href {\doibase
		10.1103/PhysRevLett.118.087004} {\bibfield  {journal} {\bibinfo  {journal}
			{Phys. Rev. Lett.}\ }\textbf {\bibinfo {volume} {118}},\ \bibinfo {pages}
		{087004} (\bibinfo {year} {2017})}\BibitemShut {NoStop}%
	\bibitem [{\citenamefont {Pinon}\ \emph {et~al.}(2020)\citenamefont {Pinon},
		\citenamefont {Kaladzhyan},\ and\ \citenamefont {Bena}}]{Pinon2020}%
	\BibitemOpen
	\bibfield  {author} {\bibinfo {author} {\bibfnamefont {S.}~\bibnamefont
			{Pinon}}, \bibinfo {author} {\bibfnamefont {V.}~\bibnamefont {Kaladzhyan}}, \
		and\ \bibinfo {author} {\bibfnamefont {C.}~\bibnamefont {Bena}},\ }\bibfield
	{title} {\enquote {\bibinfo {title} {Surface green's functions and boundary
				modes using impurities: Weyl semimetals and topological insulators},}\ }\href
	{\doibase 10.1103/PhysRevB.101.115405} {\bibfield  {journal} {\bibinfo
			{journal} {Phys. Rev. B}\ }\textbf {\bibinfo {volume} {101}},\ \bibinfo
		{pages} {115405} (\bibinfo {year} {2020})}\BibitemShut {NoStop}%
	\bibitem [{\citenamefont {Rüßmann}\ \emph {et~al.}(2021)\citenamefont
		{Rüßmann}, \citenamefont {Mavropoulos},\ and\ \citenamefont
		{Bl{\"u}gel}}]{Russmann2021}%
	\BibitemOpen
	\bibfield  {author} {\bibinfo {author} {\bibfnamefont {P.}~\bibnamefont
			{Rüßmann}}, \bibinfo {author} {\bibfnamefont {P.}~\bibnamefont
			{Mavropoulos}}, \ and\ \bibinfo {author} {\bibfnamefont {S.}~\bibnamefont
			{Bl{\"u}gel}},\ }\bibfield  {title} {\enquote {\bibinfo {title} {{Ab Initio
					Theory of Fourier-Transformed Quasiparticle Interference Maps and Application
					to the Topological Insulator Bi2Te3}},}\ }\href {\doibase
		https://doi.org/10.1002/pssb.202000031} {\bibfield  {journal} {\bibinfo
			{journal} {physica status solidi (b)}\ }\textbf {\bibinfo {volume} {258}},\
		\bibinfo {pages} {2000031} (\bibinfo {year} {2021})}\BibitemShut {NoStop}%
	\bibitem [{\citenamefont {Derry}\ \emph {et~al.}(2015)\citenamefont {Derry},
		\citenamefont {Mitchell},\ and\ \citenamefont {Logan}}]{Derry2015}%
	\BibitemOpen
	\bibfield  {author} {\bibinfo {author} {\bibfnamefont {P.~G.}\ \bibnamefont
			{Derry}}, \bibinfo {author} {\bibfnamefont {A.~K.}\ \bibnamefont {Mitchell}},
		\ and\ \bibinfo {author} {\bibfnamefont {D.~E.}\ \bibnamefont {Logan}},\
	}\bibfield  {title} {\enquote {\bibinfo {title} {Quasiparticle interference
				from magnetic impurities},}\ }\href {\doibase 10.1103/PhysRevB.92.035126}
	{\bibfield  {journal} {\bibinfo  {journal} {Phys. Rev. B}\ }\textbf {\bibinfo
			{volume} {92}},\ \bibinfo {pages} {035126} (\bibinfo {year}
		{2015})}\BibitemShut {NoStop}%
	\bibitem [{\citenamefont {Mitra}\ \emph {et~al.}(2021)\citenamefont {Mitra},
		\citenamefont {Corticelli}, \citenamefont {Ribeiro},\ and\ \citenamefont
		{McClarty}}]{mitra_magnon_2021}%
	\BibitemOpen
	\bibfield  {author} {\bibinfo {author} {\bibfnamefont {A.}~\bibnamefont
			{Mitra}}, \bibinfo {author} {\bibfnamefont {A.}~\bibnamefont {Corticelli}},
		\bibinfo {author} {\bibfnamefont {P.}~\bibnamefont {Ribeiro}}, \ and\
		\bibinfo {author} {\bibfnamefont {P.~A.}\ \bibnamefont {McClarty}},\
	}\bibfield  {title} {\enquote {\bibinfo {title} {{Magnon Interference
					Tunneling Spectroscopy as a Probe of 2D Magnetism}},}\ }\href
	{http://arxiv.org/abs/2110.02662} {\bibfield  {journal} {\bibinfo  {journal}
			{arXiv:2110.02662}\ } (\bibinfo {year} {2021})}\BibitemShut {NoStop}%
	\bibitem [{\citenamefont {Rhodes}\ \emph {et~al.}(2022)\citenamefont {Rhodes},
		\citenamefont {Osmolska}, \citenamefont {Marques},\ and\ \citenamefont
		{Wahl}}]{3dqpidata}%
	\BibitemOpen
	\bibfield  {author} {\bibinfo {author} {\bibfnamefont {L.~C.}\ \bibnamefont
			{Rhodes}}, \bibinfo {author} {\bibfnamefont {W.}~\bibnamefont {Osmolska}},
		\bibinfo {author} {\bibfnamefont {C.~A.}\ \bibnamefont {Marques}}, \ and\
		\bibinfo {author} {\bibfnamefont {P.}~\bibnamefont {Wahl}},\ }\href
	{https://dx.doi.org/<to be inserted on acceptance>} {} (\bibinfo {year}
	{2022}),\ \bibinfo {note} {{U}nderpinning data for 'On the nature of
		quasiparticle interference in three dimensions'}\BibitemShut {NoStop}%
	\bibitem [{\citenamefont {Pizzi}\ \emph {et~al.}(2017)\citenamefont {Pizzi},
		\citenamefont {Andreussi}, \citenamefont {Brumme}, \citenamefont {Bunau},
		\citenamefont {Nardelli}, \citenamefont {Calandra}, \citenamefont {Car},
		\citenamefont {Cavazzoni}, \citenamefont {Ceresoli}, \citenamefont
		{Cococcioni}, \citenamefont {Colonna}, \citenamefont {Carnimeo},
		\citenamefont {Corso}, \citenamefont {de~Gironcoli}, \citenamefont {Delugas},
		\citenamefont {DiStasio}, \citenamefont {Ferretti}, \citenamefont {Floris},
		\citenamefont {Fratesi}, \citenamefont {Fugallo}, \citenamefont {Gebauer},
		\citenamefont {Gerstmann}, \citenamefont {Giustino}, \citenamefont {Gorni},
		\citenamefont {Jia}, \citenamefont {Kawamura}, \citenamefont {Ko},
		\citenamefont {Kokalj}, \citenamefont {Kü{\c{c}}ükbenli}, \citenamefont
		{Lazzeri}, \citenamefont {Marsili}, \citenamefont {Marzari}, \citenamefont
		{Mauri}, \citenamefont {Nguyen}, \citenamefont {Nguyen}, \citenamefont {de-la
			Roza}, \citenamefont {Paulatto}, \citenamefont {Ponc{\'{e}}}, \citenamefont
		{Rocca}, \citenamefont {Sabatini}, \citenamefont {Santra}, \citenamefont
		{Schlipf}, \citenamefont {Seitsonen}, \citenamefont {Smogunov}, \citenamefont
		{Timrov}, \citenamefont {Thonhauser}, \citenamefont {Umari}, \citenamefont
		{Vast}, \citenamefont {Wu},\ and\ \citenamefont {Baroni}}]{Giannozzi_2017}%
	\BibitemOpen
	\bibfield  {author} {\bibinfo {author} {\bibfnamefont {G.}~\bibnamefont
			{Pizzi}}, \bibinfo {author} {\bibfnamefont {O.}~\bibnamefont {Andreussi}},
		\bibinfo {author} {\bibfnamefont {T.}~\bibnamefont {Brumme}}, \bibinfo
		{author} {\bibfnamefont {O.}~\bibnamefont {Bunau}}, \bibinfo {author}
		{\bibfnamefont {M.~Buongiorno}\ \bibnamefont {Nardelli}}, \bibinfo {author}
		{\bibfnamefont {M.}~\bibnamefont {Calandra}}, \bibinfo {author}
		{\bibfnamefont {R.}~\bibnamefont {Car}}, \bibinfo {author} {\bibfnamefont
			{C.}~\bibnamefont {Cavazzoni}}, \bibinfo {author} {\bibfnamefont
			{D.}~\bibnamefont {Ceresoli}}, \bibinfo {author} {\bibfnamefont
			{M.}~\bibnamefont {Cococcioni}}, \bibinfo {author} {\bibfnamefont
			{N.}~\bibnamefont {Colonna}}, \bibinfo {author} {\bibfnamefont
			{I.}~\bibnamefont {Carnimeo}}, \bibinfo {author} {\bibfnamefont {A.~Dal}\
			\bibnamefont {Corso}}, \bibinfo {author} {\bibfnamefont {S.}~\bibnamefont
			{de~Gironcoli}}, \bibinfo {author} {\bibfnamefont {P.}~\bibnamefont
			{Delugas}}, \bibinfo {author} {\bibfnamefont {R.~A.}\ \bibnamefont
			{DiStasio}}, \bibinfo {author} {\bibfnamefont {A.}~\bibnamefont {Ferretti}},
		\bibinfo {author} {\bibfnamefont {A.}~\bibnamefont {Floris}}, \bibinfo
		{author} {\bibfnamefont {G.}~\bibnamefont {Fratesi}}, \bibinfo {author}
		{\bibfnamefont {G.}~\bibnamefont {Fugallo}}, \bibinfo {author} {\bibfnamefont
			{R.}~\bibnamefont {Gebauer}}, \bibinfo {author} {\bibfnamefont
			{U.}~\bibnamefont {Gerstmann}}, \bibinfo {author} {\bibfnamefont
			{F.}~\bibnamefont {Giustino}}, \bibinfo {author} {\bibfnamefont
			{T.}~\bibnamefont {Gorni}}, \bibinfo {author} {\bibfnamefont
			{J.}~\bibnamefont {Jia}}, \bibinfo {author} {\bibfnamefont {M.}~\bibnamefont
			{Kawamura}}, \bibinfo {author} {\bibfnamefont {H-Y.}\ \bibnamefont {Ko}},
		\bibinfo {author} {\bibfnamefont {A.}~\bibnamefont {Kokalj}}, \bibinfo
		{author} {\bibfnamefont {E.}~\bibnamefont {Kü{\c{c}}ükbenli}}, \bibinfo
		{author} {\bibfnamefont {M.}~\bibnamefont {Lazzeri}}, \bibinfo {author}
		{\bibfnamefont {M.}~\bibnamefont {Marsili}}, \bibinfo {author} {\bibfnamefont
			{N.}~\bibnamefont {Marzari}}, \bibinfo {author} {\bibfnamefont
			{F.}~\bibnamefont {Mauri}}, \bibinfo {author} {\bibfnamefont {N.~L.}\
			\bibnamefont {Nguyen}}, \bibinfo {author} {\bibfnamefont {H-V.}\ \bibnamefont
			{Nguyen}}, \bibinfo {author} {\bibfnamefont {A.~Otero}\ \bibnamefont {de-la
				Roza}}, \bibinfo {author} {\bibfnamefont {L.}~\bibnamefont {Paulatto}},
		\bibinfo {author} {\bibfnamefont {S.}~\bibnamefont {Ponc{\'{e}}}}, \bibinfo
		{author} {\bibfnamefont {D.}~\bibnamefont {Rocca}}, \bibinfo {author}
		{\bibfnamefont {R.}~\bibnamefont {Sabatini}}, \bibinfo {author}
		{\bibfnamefont {B.}~\bibnamefont {Santra}}, \bibinfo {author} {\bibfnamefont
			{M.}~\bibnamefont {Schlipf}}, \bibinfo {author} {\bibfnamefont {A.~P.}\
			\bibnamefont {Seitsonen}}, \bibinfo {author} {\bibfnamefont {A.}~\bibnamefont
			{Smogunov}}, \bibinfo {author} {\bibfnamefont {I.}~\bibnamefont {Timrov}},
		\bibinfo {author} {\bibfnamefont {T.}~\bibnamefont {Thonhauser}}, \bibinfo
		{author} {\bibfnamefont {P.}~\bibnamefont {Umari}}, \bibinfo {author}
		{\bibfnamefont {N.}~\bibnamefont {Vast}}, \bibinfo {author} {\bibfnamefont
			{X.}~\bibnamefont {Wu}}, \ and\ \bibinfo {author} {\bibfnamefont
			{S.}~\bibnamefont {Baroni}},\ }\bibfield  {title} {\enquote {\bibinfo {title}
			{Advanced capabilities for materials modelling with quantum {ESPRESSO}},}\
	}\href {\doibase 10.1088/1361-648x/aa8f79} {\bibfield  {journal} {\bibinfo
			{journal} {Journal of Physics: Condensed Matter}\ }\textbf {\bibinfo {volume}
			{29}},\ \bibinfo {pages} {465901} (\bibinfo {year} {2017})}\BibitemShut
	{NoStop}%
	\bibitem [{\citenamefont {Pizzi}\ \emph {et~al.}(2020)\citenamefont {Pizzi},
		\citenamefont {Vitale}, \citenamefont {Arita}, \citenamefont {Bl{\"u}gel},
		\citenamefont {Freimuth}, \citenamefont {G{\'{e}}ranton}, \citenamefont
		{Gibertini}, \citenamefont {Gresch}, \citenamefont {Johnson}, \citenamefont
		{Koretsune}, \citenamefont {Iba{\~{n}}ez-Azpiroz}, \citenamefont {Lee},
		\citenamefont {Lihm}, \citenamefont {Marchand}, \citenamefont {Marrazzo},
		\citenamefont {Mokrousov}, \citenamefont {Mustafa}, \citenamefont {Nohara},
		\citenamefont {Nomura}, \citenamefont {Paulatto}, \citenamefont
		{Ponc{\'{e}}}, \citenamefont {Ponweiser}, \citenamefont {Qiao}, \citenamefont
		{Th{\"o}le}, \citenamefont {Tsirkin}, \citenamefont {Wierzbowska},
		\citenamefont {Marzari}, \citenamefont {Vanderbilt}, \citenamefont {Souza},
		\citenamefont {Mostofi},\ and\ \citenamefont {Yates}}]{Pizzi2020}%
	\BibitemOpen
	\bibfield  {author} {\bibinfo {author} {\bibfnamefont {G.}~\bibnamefont
			{Pizzi}}, \bibinfo {author} {\bibfnamefont {V.}~\bibnamefont {Vitale}},
		\bibinfo {author} {\bibfnamefont {R.}~\bibnamefont {Arita}}, \bibinfo
		{author} {\bibfnamefont {S.}~\bibnamefont {Bl{\"u}gel}}, \bibinfo {author}
		{\bibfnamefont {F.}~\bibnamefont {Freimuth}}, \bibinfo {author}
		{\bibfnamefont {G.}~\bibnamefont {G{\'{e}}ranton}}, \bibinfo {author}
		{\bibfnamefont {M.}~\bibnamefont {Gibertini}}, \bibinfo {author}
		{\bibfnamefont {D.}~\bibnamefont {Gresch}}, \bibinfo {author} {\bibfnamefont
			{Ch.}\ \bibnamefont {Johnson}}, \bibinfo {author} {\bibfnamefont {Ta.}\
			\bibnamefont {Koretsune}}, \bibinfo {author} {\bibfnamefont {J.}~\bibnamefont
			{Iba{\~{n}}ez-Azpiroz}}, \bibinfo {author} {\bibfnamefont {H.}~\bibnamefont
			{Lee}}, \bibinfo {author} {\bibfnamefont {J-M.}\ \bibnamefont {Lihm}},
		\bibinfo {author} {\bibfnamefont {D.}~\bibnamefont {Marchand}}, \bibinfo
		{author} {\bibfnamefont {A.}~\bibnamefont {Marrazzo}}, \bibinfo {author}
		{\bibfnamefont {Y.}~\bibnamefont {Mokrousov}}, \bibinfo {author}
		{\bibfnamefont {J.~I.}\ \bibnamefont {Mustafa}}, \bibinfo {author}
		{\bibfnamefont {Y.}~\bibnamefont {Nohara}}, \bibinfo {author} {\bibfnamefont
			{Y.}~\bibnamefont {Nomura}}, \bibinfo {author} {\bibfnamefont
			{L.}~\bibnamefont {Paulatto}}, \bibinfo {author} {\bibfnamefont
			{S.}~\bibnamefont {Ponc{\'{e}}}}, \bibinfo {author} {\bibfnamefont
			{T.}~\bibnamefont {Ponweiser}}, \bibinfo {author} {\bibfnamefont
			{J.}~\bibnamefont {Qiao}}, \bibinfo {author} {\bibfnamefont {F.}~\bibnamefont
			{Th{\"o}le}}, \bibinfo {author} {\bibfnamefont {S.~S.}\ \bibnamefont
			{Tsirkin}}, \bibinfo {author} {\bibfnamefont {M.}~\bibnamefont
			{Wierzbowska}}, \bibinfo {author} {\bibfnamefont {N.}~\bibnamefont
			{Marzari}}, \bibinfo {author} {\bibfnamefont {D.}~\bibnamefont {Vanderbilt}},
		\bibinfo {author} {\bibfnamefont {I.}~\bibnamefont {Souza}}, \bibinfo
		{author} {\bibfnamefont {A.~A.}\ \bibnamefont {Mostofi}}, \ and\ \bibinfo
		{author} {\bibfnamefont {J.~R.}\ \bibnamefont {Yates}},\ }\bibfield  {title}
	{\enquote {\bibinfo {title} {{Wannier90 as a community code: new features and
					applications}},}\ }\href {\doibase 10.1088/1361-648x/ab51ff} {\bibfield
		{journal} {\bibinfo  {journal} {Journal of Physics: Condensed Matter}\
		}\textbf {\bibinfo {volume} {32}},\ \bibinfo {pages} {165902} (\bibinfo
		{year} {2020})}\BibitemShut {NoStop}%
\end{thebibliography}
\end{document}